%% file: EDM_LOI.tex
\documentclass[11pt]{article}

\usepackage{subfigure}
\usepackage{color,amsmath,amssymb,amsfonts,amsbsy}
\usepackage{bbm}
\usepackage{bm}
\usepackage[dvipsnames]{xcolor}
\usepackage[final]{graphics}
\usepackage{epsfig}
\usepackage{sidecap}
\usepackage{hyperref}
\usepackage{lipsum}
\usepackage{graphicx}
\usepackage{color}
\usepackage{subfigure}
\usepackage{bm}
\usepackage{sidecap, caption}
\usepackage{relsize}
\usepackage{listings}
\usepackage{tabularx}
\usepackage{listings}
\usepackage{inconsolata}
\usepackage{caption}
\usepackage{placeins} 
\usepackage{cancel}
\usepackage{orcidlink}
\usepackage{cite}
\usepackage{siunitx}
\DeclareSIUnit{\rad}{rad}

\newcommand*{\affaddr}[1]{#1} 
\newcommand*{\affmark}[1][*]{\textsuperscript{#1}}

\usepackage{ulem}

\def\empile#1\over#2{\mathrel{\mathop{\kern 0pt#1}\limits_{#2}}}

\def\beq{\begin{equation}}
\def\eeq{\end{equation}}
\def\bea{\begin{eqnarray}}
\def\eea{\end{eqnarray}}

\newcommand{\Lb}{\left(}
\newcommand{\Rb}{\right)}
\def\p{{\boldsymbol p}}

\def\d3p{\frac{d^3\p}{(2\pi)^3}E_\p}

\catcode`\@=11

\newcount\@tempcntc
\def\@citex[#1]#2{\if@filesw\immediate\write\@auxout{\string\citation{#2}}\fi
  \@tempcnta\z@\@tempcntb\m@ne\def\@citea{}\@cite{%
        \@for\@citeb:=#2\do%
    {\@ifundefined{b@\@citeb}%
        {\@citeo\@tempcntb\m@ne\@citea%
                \def\@citea{,\penalty\@m\ }{\bf ?}\@warning%
                {Citation `\@citeb' on page \thepage \space undefined}}%
        {\setbox\z@\hbox{\global\@tempcntc0\csname b@\@citeb\endcsname\relax}
     \ifnum\@tempcntc=\z@ \@citeo\@tempcntb\m@ne%
       \@citea\def\@citea{,\penalty\@m}%
       \hbox{\csname b@\@citeb\endcsname}%
     \else%
      \advance\@tempcntb\@ne%
      \ifnum\@tempcntb=\@tempcntc%
      \else\advance\@tempcntb\m@ne\@citeo%
      \@tempcnta\@tempcntc\@tempcntb\@tempcntc\fi\fi}}\@citeo}{#1}}%

\def\@citeo{\ifnum\@tempcnta>\@tempcntb\else\@citea
  \def\@citea{,\penalty\@m}%
  \ifnum\@tempcnta=\@tempcntb\the\@tempcnta\else
   {\advance\@tempcnta\@ne\ifnum\@tempcnta=\@tempcntb \else
\def\@citea{--}\fi
    \advance\@tempcnta\m@ne\the\@tempcnta\@citea\the\@tempcntb}\fi\fi}

\catcode`\@=12

\begin{document}

\input{title.tex}

\date{}
\begin{abstract}
A breakthrough in fundamental physics experiments measuring particle spin precession may happen if spin-transparent storage rings become adopted tools for such experiments. 
We present a new design of highly specialized table-sized storage rings, which use low-energy polarized electron beams and Mott polarimetry. Based on the spin transparency \textit{ansatz}, 
the spin precession stemming from the magnetic dipole moment is canceled at any beam energy after an electron's turn along the periodic orbit in the ring. Meanwhile, a spin precession 
induced by the fundamental physics of interest, e.g., the electron's permanent electric dipole moment (EDM) and/or ultralight-dark-matter-mediated forces such as axions, will accumulate. 
However, capitalizing on such types of rings is not only desirable for measurements of EDMs and axion searches relevant to $CP$ violation and matter-antimatter asymmetry in the Universe, 
but may also find very promising applications in quantum computing.
\end{abstract}

\tableofcontents

\clearpage
\section{Introduction}
\label{sec:intro}

In this paper we mostly use the materials from Refs.~\cite{Suleiman:2023whz,Suleiman:2021whz,Suleiman:2024hvh}. In particular, in Secs. \ref{sec:concept} and \ref{sec:prec_edm}, we follow the 
line of reasoning from \cite{Suleiman:2023whz,Suleiman:2021whz} to describe a high-precision measurement of the electron's permanent electric dipole moment (eEDM) with a spin-transparent (ST) 
storage ring. Section~\ref{sec:search_axion} rests upon similar concepts from \cite{Suleiman:2023whz,Suleiman:2024hvh} to outline the search of ultralight axion dark matter with a bit different 
type of ST ring. In Sec.~\ref{sec:pos_edm}, we discuss some prospects for the positron's permanent electric dipole moment measurement. Meanwhile, in this introduction we give some details related 
to EDMs and axions in general, along with motivating this study. Those ST rings and technology can also have another exciting application, which is the quantum computing \cite{Suleiman:2024hvh,Suleiman:2026zxn}.

\subsection{Electron's permanent electric dipole moment}
\label{sec:edm}

Permanent electric dipole moments of fundamental particles, such as the electron, proton and neutron, are signatures of time-reversal ($T$) symmetry violation -- equivalent to violations of 
combined charge and parity ($CP$) symmetry \cite{Khriplovich:1997ga}. $CP$ symmetry is broken in the Standard Model (SM) but only in the quark sector, where  the coupling to leptons is weak and, 
for example, the predicted eEDM is several orders of magnitude below the current experimental sensitivity \cite{Yamaguchi:2020eub,Ema:2022yra}. Since the eEDM tests the $CP$ violation via leptons, 
these tests are often complementary to nuclear searches (such as nEDM), which constrains $CP$-violating interactions between quarks and gluons. Therefore, the measurement of the permanent eEDM 
is intimately connected to nuclear and particle physics, particularly through the $CP$ violation in high-energy processes and via the use of heavy atoms/molecules as enhanced laboratories. 
In principle, particle EDM precision measurements allow for improved constraints on theoretical $CP$ violation within and beyond the SM. 

The SM itself cannot explain the puzzle related to the dominance of matter over anti-matter in our Universe \cite{Dine:2003ax}. This imbalance points out undiscovered physics, which violates $CP$ 
symmetry \cite{Sakharov:1967dj,Gavela:1993ts}. So, explaining this matter and anti-matter asymmetry in the Universe requires additional $CP$ violation, beyond what is present in the weak interactions 
of the SM \cite{Gavela:1993ts}. Some SM extensions seek to explain the imbalance by predicting the existence of new particles \cite{Engel:2013lsa}. Vacuum fluctuations of the fields associated 
with such new particles can interact with known particles and imply small modifications of their properties, e.g., particles that violate $CP$ symmetry induce an EDM. Then, the size of the induced 
EDM will be dependent on the new particles' masses and their couplings to the SM particles.  Thereby, the EDM is very sensitive to physics beyond the SM and new sources of $CP$ violation (such as, 
the $\theta$-term\footnote{The $\theta$-term in the SM refers to a $CP$-violating term that is allowed in Quantum Chromodynamics (QCD). Its presence would violate the $CP$ symmetry in the strong 
interactions. Likewise, it is the absence of $CP$ violation in the strong interactions, which has led to formulation of the Strong $CP$ problem.}). It is a signal of new physics and helps explain 
the matter-antimatter imbalance puzzle in the Universe \cite{Chupp:2017rkp,Yamanaka:2016umw}.

\hskip -0.6truecm
The classical definition of the eEDM is given by the quantity
\beq
\vec{d_{e}} = \int (\vec{s} - \vec{s}_{0})\,\rho(\vec{s})\,d^{3}\vec{s} ,
\label{eq:eq_edm}
\eeq
but which is dependent on the coordinate offset $\vec{s}_{0}$ of the charge distribution. The eEDM must be collinear with the direction of the electron's magnetic moment (spin). 
It is an intrinsic property of the electron such that the potential energy is linearly related to the strength of the electric field: $U = -\vec{d_{e}}\!\cdot\!\vec{E}$. Within the SM, 
$d_{e}$ is predicted to be non-zero but very small, at most $10^{-38}~e$\,cm \cite{Pospelov:2005pr}, where $e$ stands for the elementary charge. Specifically, the discovery of a 
substantially larger $d_{e}$ would imply a violation of both $T$ invariance  and $CP$ invariance as discussed above. 
The eEDM is usually not measured in free electrons, but instead in bound and unpaired valence electrons inside atoms/molecules, where one can observe the effect of  
$-\vec{d_{e}}\!\cdot\!\vec{E}$ as a slight shift of spectral lines.

The $d_{e}$ sensitivity scales approximately with the nuclear charge cubed \cite{Alarcon:2022ero}. That is why eEDM searches are usually conducted on systems involving heavy 
atoms/molecules. The following experimental upper limits at 90\% Confidence Level (C.L.) have been put in indirect measurements of eEDM:
\begin{itemize}
\item[$\bullet$] \underline{Molecular beam} of Ytterbium monofluoride (YbF): $d_{e} <1.1 \times 10^{-27}~e$\,cm \cite{Hudson:2011zz}; \\
\underline{Molecular beam} of Thorium monoxide (ThO): $d_{e} < 8.7 \times 10^{-29}~e$\,cm \cite{ACME:2013pal}, \\
$ d_{e} < 1.1 \times 10^{-29}~e$\,cm \cite{ACME:2018yjb}; \\
\underline{Molecular beam} of Barium fluoride (BaF): expected sensitivity $d_{e} <5 \times 10^{-30}~e$\,cm (ongoing) \cite{NLeEDM:2018}. 

\item[$\bullet$] \underline{Atomic beam} of Thallium (TI): $d_{e} < 1.6 \times 10^{-27}~e$\,cm \cite{Regan:2002ta}.

\item[$\bullet$] \underline{Molecular ion trap} of Hafnium fluoride ion (HfF$^{+}$): $d_{e} < 1.3 \times 10^{-28}~e$\,cm \cite{Cairncross:2017fip}; \\
\underline{Molecular ion trap} of Hafnium fluoride ion (HfF$^{+}$): $d_{e} < 4.1 \times 10^{-30}~e$\,cm \cite{Roussy:2022cmp} \footnote{A very recent work of 
Ref.~\cite{Ardu:2025rqy} based on using the Standard Model Effective Field Theory (SMEFT) shows systematic computations of contributions to the full low-energy 
domain probed by EDM searches with a sensitivity equivalent to the eEDM precision's reach of the HfF$^{+}$ molecular ions.}. 

\item[$\bullet$] \underline{Molecular trap} of Ytterbium monohydroxide (YbOH): (ongoing) \cite{Kozyryev:2017cwq}. 
\end{itemize}

\medskip
\hskip -0.6truecm
Direct measurements of the EDM upper limits exist for the other two charged leptons:
\begin{itemize}
\item[$\bullet$] for the muon $d_{\mu} < 1.8 \times 10^{-19}~e$\,cm with 95\% C.L. \cite{Muong-2:2008ebm};

\item[$\bullet$] for the tau $d_{\tau} < 10^{-17}~e$\,cm with 95\% C.L. \cite{Belle:2021ybo};
\end{itemize}
plus for the neutron $d_{n} < 1.8 \times 10^{-26}~e$\,cm with 90\% C.L.\cite{Abel:2020pzs}.

However, to date there are no direct measurements of the electron or proton EDMs. Nevertheless, increasingly precise measurements attempt to probe new physics with higher masses 
and weaker couplings. The EDM upper limit for the  proton has been indirectly obtained based on measurements using the neutral Mercury atom (${}^{199}$Hg) \cite{Graner:2016}, where 
the pEDM limit was determined to be $d_{p} < 2 \times 10^{-25}~e$\,cm. However, there are plans for measuring $d_{p}$ in storage rings 
\cite{Anastassopoulos:2015ura,Haciomeroglu:2018nre,1865072,Omarov:2020kws}. We also refer to Ref.~\cite{Alarcon:2022ero} for a review on various EDM measurements and a feasibility 
study of such measurements in storage rings \cite{CPEDM:2019nwp}.

Meanwhile, one can carry out EDM searches with light nuclei \cite{Yamanaka:2018dwa} instead of heavy systems, which comes with a benefit since a bare light nucleus has no bound electrons, 
there is no suppression of hadronic $CP$ violation due to Schiff’s screening\footnote{According to the Schiff screening \cite{Liu:2007en}, a nuclear EDM is shielded by the surrounding electron 
cloud in a neutral atom, making it challenging to measure nuclear $CP$ violation even when the relativity in the nucleus is taken into account.}, which is associated with heavy nuclei such as 
${}^{199}$Hg. The light nuclear EDM has very small SM backgrounds to be considered. These EDMs are also expected to be accurately measured in storage rings such as the pEDM 
\cite{Khriplovich:1998zq,JEDI:2016swi}.

\subsection{Ultralight dark matter: axions}
\label{sec:axion}

The most well-known proposed solution to the Strong $CP$ problem (mentioned in Sec.~\ref{sec:edm}) is the Peccei-Quinn mechanism \cite{Peccei:1977hh,Peccei:1977ur}. The latter promotes the 
$\theta$-term to a dynamical field that relaxes to zero, predicting the existence of a new particle (axion). Let us start from the QCD Lagrangian, which contains the $\theta$-term that 
violates the $CP$ symmetry:
\beq
\mathcal{L}_{\cancel{CP}} = \theta\,\frac{\alpha_{s}}{8\pi}\,G_{\mu\nu}^{a} \tilde{G}_{a}^{\mu\nu} ,
\label{eq:eq_axion1}
\eeq
where $\alpha_{s}$ is the QCD coupling constant, $G_{\mu\nu}^{a}$ is the gluon-field strength tensor and $\tilde{G}_{a}^{\mu\nu}$ is its dual. The $\theta$-term can be understood as an angle 
defining the vacuum of the theory if one considers the QCD Lagrangian in isolation. But when embedded in the full SM Lagrangian, $\theta$ obtains a contribution from the transformation of the 
quark fields necessary to remove a common phase of all quark masses. Because of this reason, it is hard to comprehend why the $\theta$-term would be zero in the SM. Although it has no effect 
in perturbative QCD calculations and can be neglected, it has observational consequences, the most important of which is the prediction of hadronic EDMs such as that of the neutron. 
It was already mentioned in Sec.~\ref{sec:edm} that $d_{n} < 1.8 \times 10^{-26}~e$\,cm with 90\% C.L. \cite{Abel:2020pzs}, and this bound in turn imposes the restriction on $\theta$ to be 
$\theta < 0.8 \times 10^{-10}$ \cite{Irastorza:2021tdu}. The core part of the Strong $CP$ problem is the smallness of $\theta$ that is composed of different phases of disparate origin.

Axions were in fact first identified by Weinberg \cite{Weinberg:1977ma} and Wilczek \cite{Wilczek:1977pj} as the pseudo-Nambu-Goldstone bosons of the new U(1) spontaneously broken global 
symmetry (broken at some high energy scale $f_{a}$) that Peccei and Quinn had postulated, and which was called PQ (Peccei-Quinn) symmetry. The boson itself was dubbed the axion by Wilczek 
\cite{Wilczek:1991jgb}, which is currently viewed as a possible main ingredient of ultralight dark matter. Despite the fact that some later solutions to the Strong $CP$ 
problem have been proposed \cite{Pospelov:2005pr,Peccei:2006as}, the Peccei-Quinn mechanism remains the most persuasive one. In this respect, we refer to recent detailed reviews on axions 
and axion models \cite{Irastorza:2021tdu,DiLuzio:2020wdo}. One can additionally note that the axions are ubiquitous in the low-energy effective theory of many string theory compactifications 
\cite{Svrcek:2006yi,Arvanitaki:2009fg}.

Now, if we focus on the low-energy effective Lagrangian of the axion of the PQ symmetry, it includes the term
\beq
\mathcal{L} \ni \frac{\alpha_{s}}{8\pi}\,G_{\mu\nu}^{a} \tilde{G}_{a}^{\mu\nu}\,\frac{A}{f_{a}} ,
\label{eq:eq_axion2}
\eeq
where the $\theta$ angle is absorbed into a redefinition of the axion field $A$. In the absence of other $CP$-violating sources, the potential imposed on the axion field by QCD dynamics 
has a minimum at the $CP$-conserving value of $\theta = 0$. Thereby, the potential of the axion is minimized when the effective $\theta$ is zero, naturally solving the Strong $CP$ problem 
(not only rendering the initial parameter unphysical but also dynamically settling it down to zero).

\begin{SCfigure}[][hbt!]
\centering
\hspace{1.0cm}
\includegraphics[width=4.0cm]{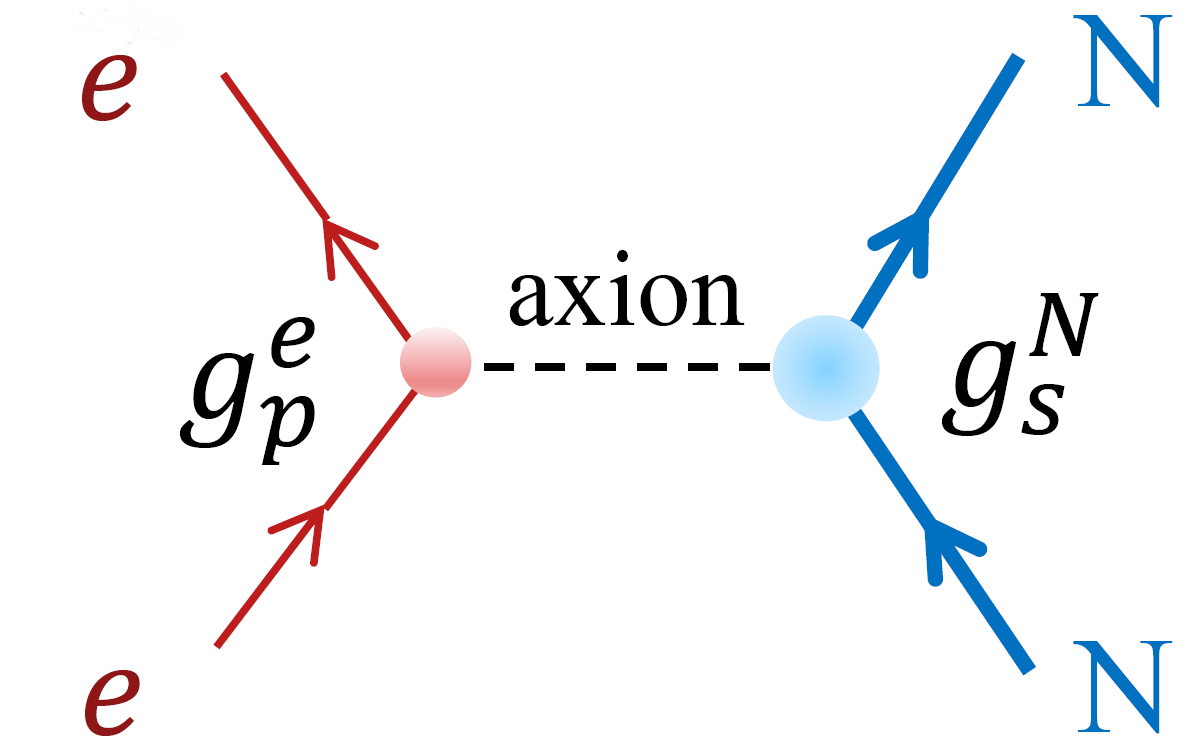}
\caption{Diagram describing the axion-mediated electron-nucleon interaction.}
\label{fig:fig_axion_diagram}
\end{SCfigure}

Some axion properties are determined by the PQ mechanism itself and are independent of the specific way it is implemented in the SM. Equation~(\ref{eq:eq_axion2}) is the defining component of 
the PQ mechanism and implies the coupling of the axion to the gluon field in particular. But the original axion models had the $f_{a}$ scale identified with the electroweak scale, and those
would lead to signatures that could not be observed in accelerators. Models with much larger $f_{a}$ scales have been proposed with values well above $10^{7}$~GeV to avoid current experimental 
constraints (see Refs.~\cite{Kim:1979if,Shifman:1979if} and \cite{Dine:1981rt,Zhitnitsky:1980tq}). Equation~(\ref{eq:eq_axion2}) also allows for mixing of the axion field with mesons, and via this
mixing the axion acquires a mass given by (see \cite{Irastorza:2021tdu})
\beq
m_{a} = 5.70(7)\,\mu{\rm eV} \Lb \frac{10^{12}~{\rm GeV}}{f_{a}} \Rb .
\label{eq:eq_axion3}
\eeq
The fact that the mass $m_{a}$ is unambiguously related to $f_{a}$ in Eq.~(\ref{eq:eq_axion3}) means that every axion coupling is also proportional to $m_{a}$, where the lowercase $a$ 
also refers to more general axion-like particle case \cite{Irastorza:2018dyq}. The axions in the current models couple to matter by mediating weak and long-range ``fifth forces" between 
macroscopic objects, acting beyond standard gravity and electromagnetism (see, e.g., a basic leading-order diagram in Fig.~\ref{fig:fig_axion_diagram}).
\FloatBarrier

The axions attribute a model-independent coupling to nucleons, stemming from the mixing with the mesons. The nucleon couplings play an important role in some ``stellar" scenarios, and consequently are 
appropriate to be used in astrophysics-based experiments to constrain axion models. The axions do not couple to electrons in a model-independent way. However, specific models may exhibit this coupling at 
tree level. Another feature of axion-meson mixing is a model-independent coupling to photons. In this case, that coupling is the source of the axion-to-photon oscillation/conversion in the background 
of an electromagnetic field, which is a mechanism sitting at the basis of axion phenomenology. 

\begin{figure}[h!]
\centering
\hspace{0.0cm}
\includegraphics[width=11.0cm]{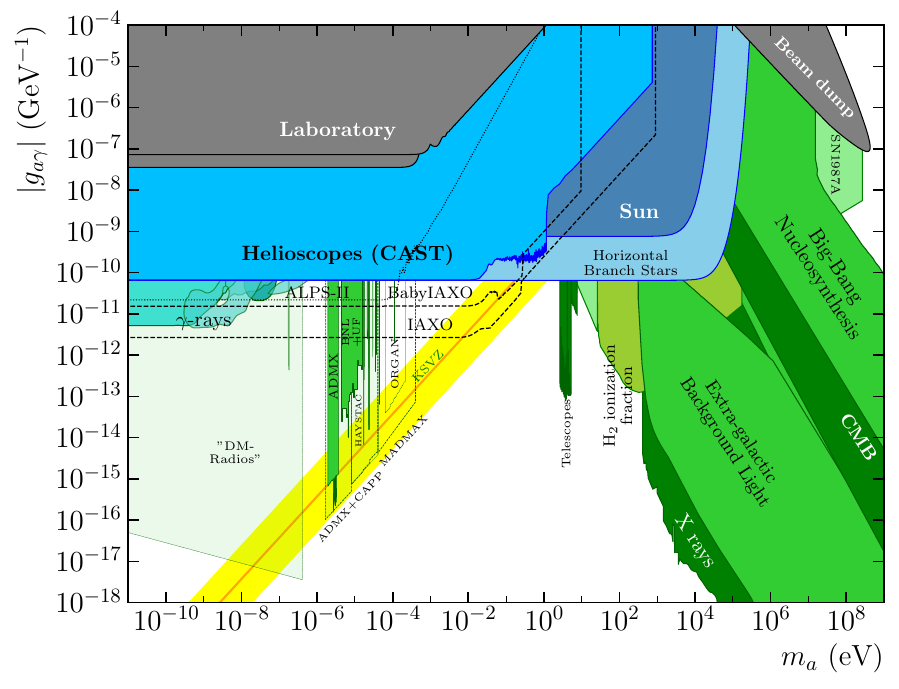}
\caption{Overall survey of current limits (dark/solid areas) and future prospects (semi-transparent areas or dashed lines) in the $g_{a\gamma}$-$m_a$ parameter plane.
This figure is from Ref.~\cite{Irastorza:2021tdu}. See its Secs.~2, 6, and 8 for more details, and Ref.~\cite{Irastorza:2018dyq} for details on the various lines.}
\label{fig:fig_panorama}
\end{figure}

The photon interaction also allows for the decay of axions into two photons. The axion-photon coupling is a necessary outcome of the PQ mechanism. Due to the importance of the axion-photon interaction 
in many of the axion experimental-detection strategies, the $(g_{a\gamma}, m_{a})$ parameter space shown in Fig.~\ref{fig:fig_panorama} remains the main arena for presenting axion results, 
observational limits, and experimental sensitivities.

Many axion models have been studied and built in the literature, and in recent years there have been intense model-building efforts in the axion phenomenology community. We again wish to mention
Refs.~\cite{Irastorza:2021tdu,DiLuzio:2020wdo} and references therein for more details and examples.

\subsection{Our motivation}
\label{sec:motiv}

The motivation of our proposed measurements are described as follows, whereas the Executive Summary of this paper has additional details.

\vspace{-0.25cm}
\paragraph{Electron EDM measurement.} 
In a specifically-designed particle mini-accelerator compact ST storage ring, based on the so-called spin transparency ansatz, the electron spin precession stemming from the magnetic dipole moment (MDM) 
will be canceled at any beam energy after the electron’s turn along the periodic orbit in the ring. Meanwhile, the spin precession induced by the fundamental physics of interest will be accumulated, e.g., 
induced by the permanent eEDM. Any spin direction repeats itself after such a periodic turn in this type of a storage ring, which is also called Figure-8 ring\footnote{For the original ideas on the 
Figure-8 ring, see Refs.~\cite{Derbenev:1975,Derbenev:2024jyn}.} as shown in Fig.~\ref{fig:fig_Figure8}, where the global spin precession is zero independent of particle energy (spin-echo effect):
\begin{SCfigure}[][hbt!]
\centering
\hspace{0.0cm}
\includegraphics[width=5.0cm]{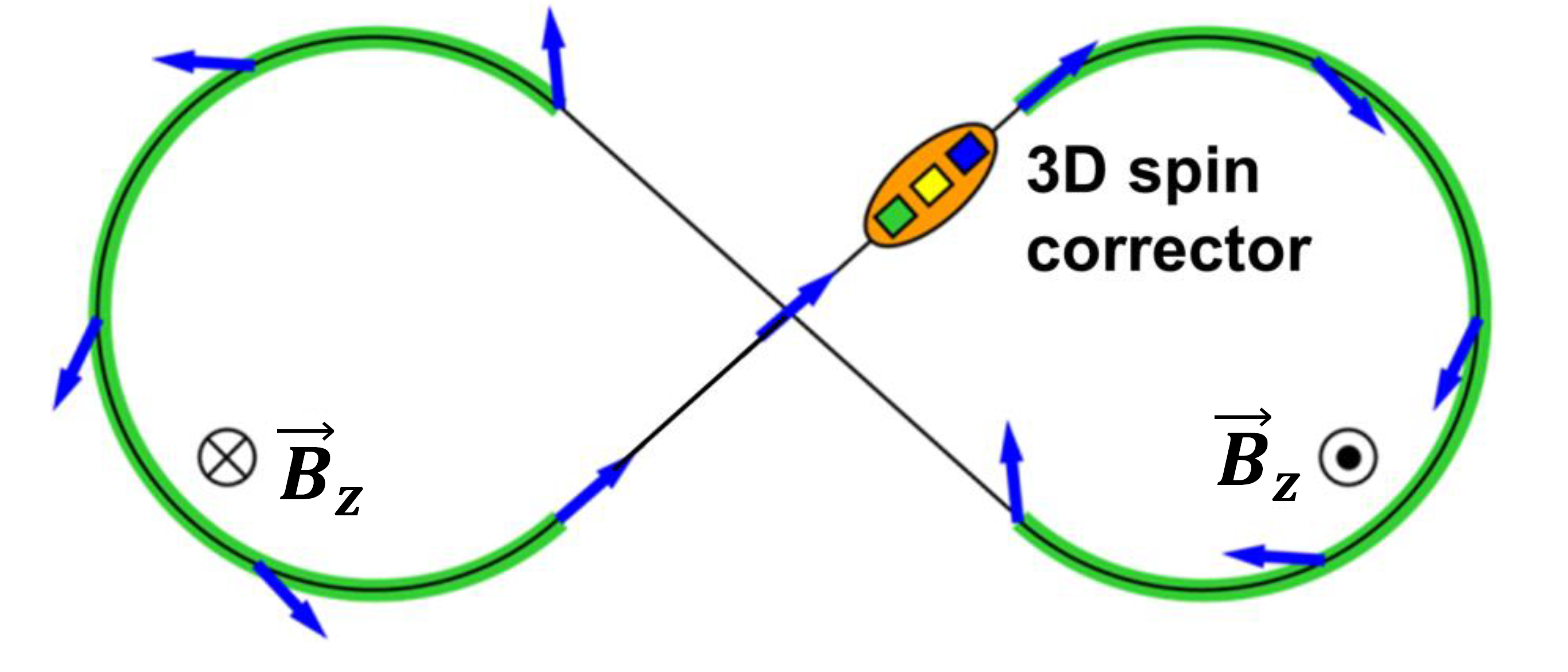}
\caption{Figure-8 ring configuration as an example of a ST storage ring, showing the most natural ST topology.}
\label{fig:fig_Figure8}
\end{SCfigure}

Thereby, this paper presents a method for measuring the eEDM in table-size storage rings with the electron's polarized beam energy range up to 1~MeV or less, based on the use of the ST Figure-8 orbit 
symmetry. We will discuss conceptual and specific ring topologies shown in Fig.~\ref{fig:fig_specific_edm}. In that case, after numerous turns, one can measure the accumulated vertical polarization 
signal caused by the eEDM; more exactly, the eEDM spin precession rate. In particular, after five years of data taking with a fully designed storage ring, the projected statistical limit for $d_{e}$ 
is about $5.8 \times 10^{-30}~e$\,cm with 90\% C.L., by also having expectations that further optimization and improvements will decrease this limit. However, the measurement with one ring 
will just be a testing ground for potential future multiple-ring measurements. From this perspective, the extractable improved precision may already be competitive with the sensitivities from 
the best and even upcoming eEDM indirect measurements discussed in Sec.~\ref{sec:edm}, like the measurement in Ref.~\cite{Roussy:2022cmp}. In the meantime, our proposed measurement with one ring would 
be the first direct eEDM measurement to be conducted ever\footnote{We wish to mention about the efforts Richard Talman has been devoted to developing methods for measuring the EDMs of the electron and 
the proton in storage rings \cite{Talman:2015pma}.} that could provide independent sanity check for indirect measurements. Similar steps may be applied to potential direct EDM measurements of the positron
as well.

\vspace{-0.25cm}
\paragraph{Axion search.}
A Figure-8 axion ST storage ring, which by its design will be similar to the eEDM ring can measure the electron spin precession frequency induced by the axion field as small as 
$0.2 \cdot 10^{-9}$~rad/s (or 0.2~nrad/s), and will be capable of providing strongest bound to surpass any existing and foreseeable search by several orders of magnitude 
(see also Fig.~\ref{fig:fig_axion_plot} in this regard). Analogously, in this case too, we will be able to measure even smaller spin precession rate by simultaneously employing 
just two axion rings. 

Except for helping solve the Strong $CP$ problem and being identified as ultralight dark matter candidate (while opening a new frontier beyond the SM), we can confidently stress that if observed, 
the axions could allow the theory community to test grand unified theories and some string theories, as shown in Refs.~\cite{Agrawal:2022lsp,Agrawal:2024ejr,Reig:2025dqb}.

\vspace{-0.25cm}
\paragraph{Technology development.}
Lastly, our motivation is also anchored on a clear vision that there will be technology development benefits; namely, (i) demonstration of a fully working two-energy ST ring with electron beams, 
(ii) demonstration of a path towards quantum computer applications \cite{Suleiman:2024hvh,Suleiman:2026zxn}.

\bigskip
\section{Spin-transparent EDM storage ring concept and design for low-energy polarized electron beams}
\label{sec:concept}

The development of an EDM measurement concept using polarized charged particle beams stored in ST rings requires a careful consideration of many technical issues associated with 
long-term beam polarization stability and lifetime. Therefore, one must necessarily formulate pertinent requirements on the beam intensity and quality. Other difficulties are related 
to polarization measurement and control. But needless to say that we have built upon the experience and innovations gathered by other research groups planning to measure pEDM 
with storage rings~\cite{Anastassopoulos:2015ura,Haciomeroglu:2018nre,CPEDM:2019nwp}.

\subsection{Spin precession}
\label{sec:precess}

Any EDM measurement is based on measuring the spin precession rate in an electric field in a particle rest frame:
\begin{equation}
  \frac{d{\vec S}}{d\tau} =  {\vec \mu} \times {\vec B}_{\rm rest} +  {\vec d} \times {\vec E}_{\rm rest} , 
  \label{eq:T-BMT_restFrame}  
\end{equation}
where the MDM and EDM are defined as ${\vec \mu}=(G+1)(q/mc){\vec S}$ and ${\vec d}=(\eta/2) (q/mc) {\vec S}$, respectively, along with $q$ and $m$ that are the particle 
charge and mass, $c$ is the speed of light, $G$ is the magnetic anomaly, and $\eta$ is the electric dipole factor. The symbol ${\vec S}$ denotes the spin as a vector in the same particle rest frame.

Nonetheless, for charged particles such a measurement cannot be carried out while a given particle is kept at rest. The reason is because the electric field leads to the particle's acceleration. 
Thus, in order to be able to apply the electric field and keep the charged particle trapped, one can use, e.g., a storage ring. For a charged particle moving in transverse $\vec{E}_{\perp}$ and 
longitudinal $\vec{E}_{||}$ electric fields as well as in a magnetic $\vec{B}$ field, there is the following generalized equation of spin precession~\cite{Fukuyama:2013ioa} given in the lab's 
Cartesian frame, called Thomas-Bargmann-Michel-Telegdie (or, Thomas-BMT) equation:
\begin{equation}
\frac{d{\vec S}}{dt} = \left( {\vec \omega}_{\rm MDM} + {\vec \omega}_{\rm EDM} \right) \times {\vec S} , 
\label{eq:eq:T-BMT_labFrame}  
\end{equation}
\noindent with
\begin{equation}
{\vec \omega}_{\rm MDM} = -\frac{q}{mc} \left( \frac{1+G}{\gamma}\,{\vec B}_{\rm rest} + \frac{1}{\gamma+1}\,{\vec \beta} \times {\vec E}_{\rm rest} \right) ,~~~~~~~
{\vec \omega}_{\rm EDM} = -\frac{\eta}{2} \frac{q}{mc} {\vec E}_{\rm rest} ,
\label{eq:eq:T-BMT_restFrame-MDM}  
\end{equation}
where ${\vec v} \equiv {\vec \beta} c$ and $\gamma$ are the particle velocity and Lorentz energy factor, respectively. The term in Eq.~(\ref{eq:eq:T-BMT_restFrame-MDM}) proportional 
to ${\vec \beta} \times {\vec E}_{\rm rest}$ emerges due to a relativistic transformation of the spin direction with the velocity rotation (Thomas precession~\cite{Thomas:1927yu}). 
Then one can obtain the following expressions in terms of electric and magnetic fields in the lab's frame~\cite{Fukuyama:2013ioa}: 
\begin{equation}
{\vec \omega}_{\rm MDM} = -\frac{q}{mc} \left[ \left( \frac{1}{\gamma} + G \right) {\vec B}_\bot + \frac{1}{\gamma} (1+G) {\vec B_\parallel} + \left( \frac{1}{\gamma + 1} + G \right) 
{\vec E} \times {\vec \beta} \right] , 
\label{eq:T-BMT_FieldslabFrame-MDM}  
\end{equation}
\begin{equation}
{\vec \omega}_{\rm EDM} = -\frac{\eta\,q}{2mc} \left( \frac{1}{\gamma}\,{\vec E_\parallel} + {\vec E}_\bot + {\vec \beta} \times {\vec B} \right) . 
\label{eq:eq:T-BMT_FieldslabFrame-EDM}  
\end{equation}

By having the discussion of Sec.~\ref{sec:motiv} continued in a bit more detailed manner, we again indicate that the general principle of an EDM measurement in a ST storage ring relies on 
making the MDM spin rotation effectively vanish by compensating this rotation either instantaneously or over a turn while allowing for turn-by-turn accumulations of the EDM precession signal. 
The observation of a spin (polarization) rotation then will point out the presence of EDM. This type of a measurement requires account of and compensation for systematic effects associated with beam 
emittances\footnote{The beam emittance is a measure of the average spread of charged particles in the position \& momentum phase space within an accelerator to describe the beam's quality.}, 
field errors, and background magnetic fields. Strategies for canceling the MDM spin precession could be formulated for a particle motion along a closed reference design orbit, conventionally 
(but not unavoidably) in a plane, which enables the effective stacking of the EDM precession turn-by-turn around that orbit. Afterwards, a flat reference orbit can be generally designed by 
using radial electric $E_x$ and vertical magnetic $B_y$ fields. In a coordinate system linked to a given particle velocity on the design orbit (in the so-called Frenet-Serret frame), 
the MDM spin precession about the vertical axis has an angular frequency of
\begin{equation}
\omega_{y,{\rm MDM}} = -\frac{q}{mc} \left( G B_y - \frac{1 - \gamma^2 \beta^2 G}{\gamma^2 \beta} E_x \right) . 
\label{eq:T-BMT_labFrame-ME}  
\end{equation}

The ST method has been originally developed for magnetic synchrotrons. It was then extensively studied while designing polarized electron, proton, deuteron, and ${^3}$He beams for the 
Electron-Ion Collider architected to be hosted at JLab~\cite{Filatov:2020ygb,Kondratenko:2019ubq}. We developed a technique \cite{Suleiman:2023whz,Suleiman:2021whz} for eEDM measurements 
by extending this method to small-size storage rings designed to use electric fields exclusively, to allow for two counter-rotating ($CR$) electron beams $A$ and $B$ ($CRA$ and $CRB$) 
to circulate concurrently. It results in having new mechanisms for suppressing destructive effects associated with beam emittances and energy spread. In the case of electrons, we developed 
a Figure-8 all-electric ($B_y=0$) ring design with polarized beams at the energy $\lesssim1$~MeV modulated along the orbit. At this low electron beam energy, Mott polarimetry can be utilized 
with its highest efficiency.

\subsection{Two-energy storage ring conceptual and specific designs}
\label{sec:edm_ring}

So, one can treat the eEDM effect in a storage ring as a tiny signal (perturbation) on top of a relatively large MDM spin precession. We begin this section by describing the conceptual design of a 
Figure-8 ST ring configuration with two beams at different energies to be operating in various ring sections. The net bend at each of the $\gamma_{1}$ and $\gamma_{2}$ energies is still zero preserving 
the MDM suppression. Those different energy sections should be connected by static longitudinal electric field sections. These fields have no effect on the MDM spin precession but break the degeneracy 
of the eEDM spin precession. The level of the expected eEDM signal will be discussed in Sec.~\ref{sec:prec_edm}. One of the main challenges for measuring the eEDM effect in a MDM-suppressed system is 
to have a precise control of systematic effects. We must confidently differentiate a real eEDM signal from a false one caused by residual MDM effects. Perhaps, the most powerful technique for suppression 
of the systematic effects is to use two $CR$ beams in a direction-symmetric ring. Since any bending magnetic field breaks the symmetry between the $CR$ beams, we select an all-electric ST ring.

We employ static longitudinal electric field sections for the electron acceleration and deceleration. They allow for beam energy recovery unlike warm RF (radiofrequency) cavities, thus minimizing the ring's 
energy consumption. On the other hand, the static longitudinal fields do not provide beam bunching (that is carried out by RF cavities)\footnote{In the accelerator physics, the beam bunching is a process of 
clustering a particle beam longitudinally into distinct pulses or ``bunches" rather than making a continuous stream.}, which is needed for time separation of the $CR$ bunches. Time separation of the bunches 
is in turn needed for their simultaneous independent stochastic cooling\footnote{The stochastic cooling is a form of particle-beam cooling, which reduces the energy spread and angular divergence of a particle 
beam (cooling) by measuring and correcting individual particle deviations.}. The stochastic cooling is necessary to maintain the bunch intensities sufficient from statistical-accuracy point of view. The 
beam bunching will allow for simultaneous storage of high-intensity $CR$ bunches, as a critical tool for suppression of many systematic effects. Therefore, we also introduce a dedicated RF cavity to bunch 
the beams. The RF electric field of the bunching cavity is accompanied by an oscillating magnetic field, which interacts with the particle's MDM. Meanwhile, its effect averages to zero over time due to 
synchrotron oscillations.  

\begin{figure}[hbt]
\vspace{-0.65cm}
\centering
\includegraphics[width=5.85cm]{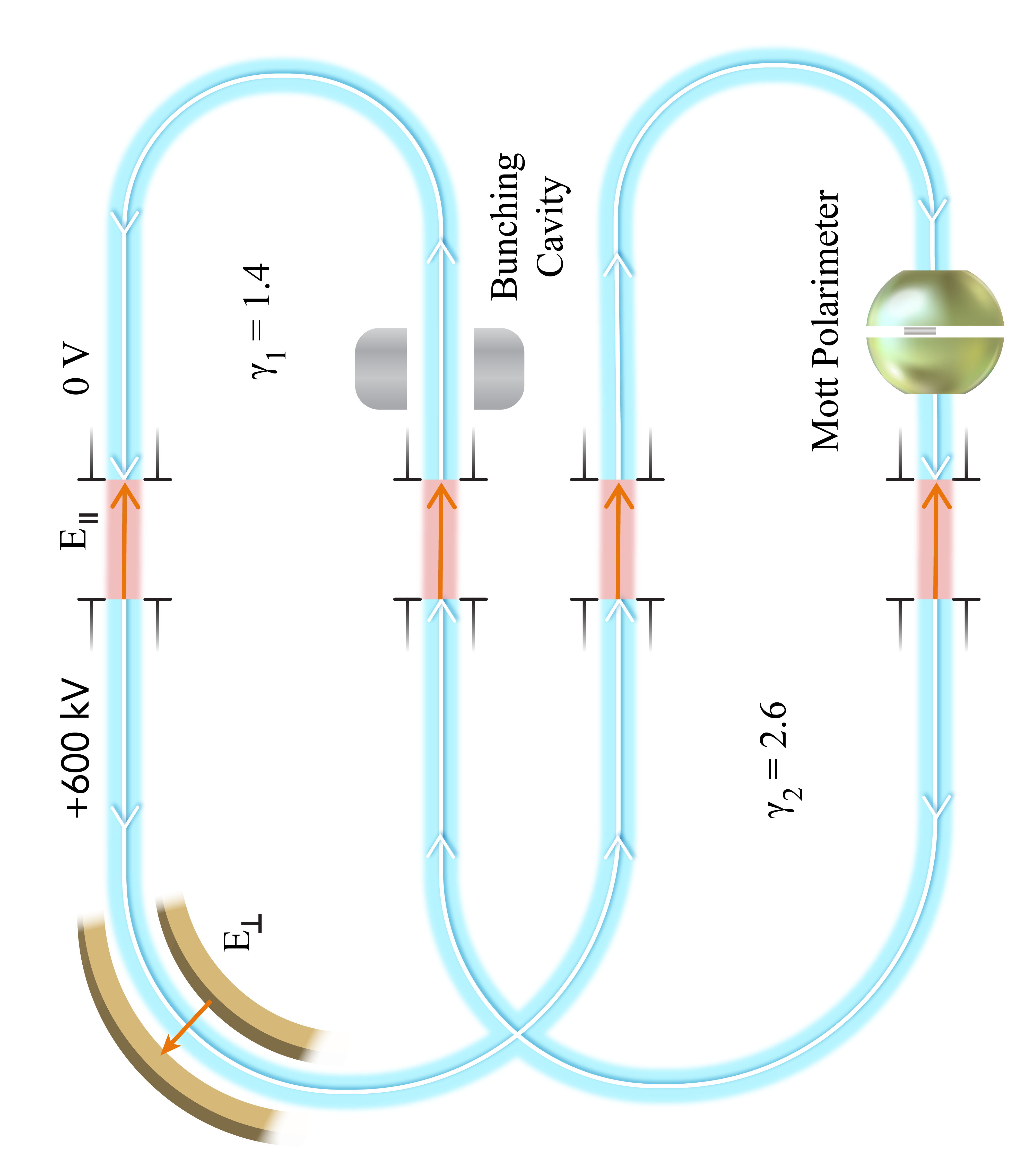}
\hspace{0.25cm}
\includegraphics[width=7.85cm]{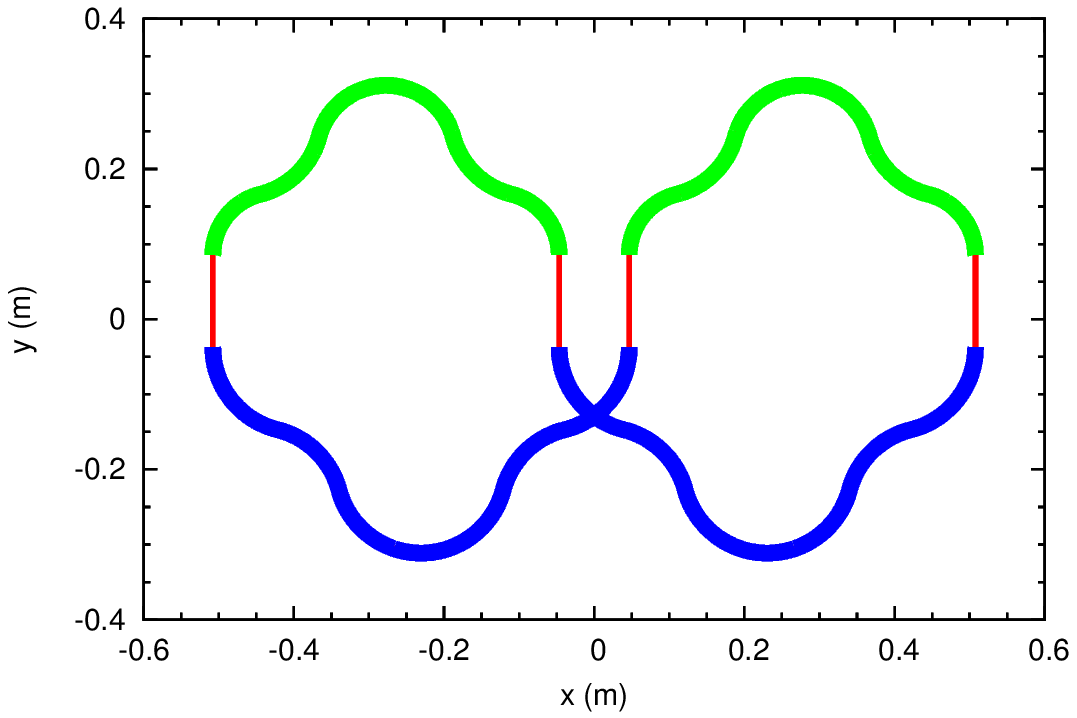}
\caption{\textbf{Left panel:} Layout of the conceptual design of a two-energy ST storage ring for the eEDM measurement (the figure is not drawn to scale). The ring uses only static electric fields ($E_\parallel$ 
and $E_\perp$) except for a single RF bunching cavity. Only one of the two $CR$ electron beams is shown. The high-energy arcs are floating at high voltage of 600~kV. The horizontal and vertical polarization 
components can be simultaneously measured using a Mott polarimeter. \textbf{Right panel:} Footprint of a similar but specific design of a two-energy ST ring for the eEDM measurement (the figure is drawn to 
scale), where the high-energy arcs are in blue and the low-energy arcs are in green, along with the longitudinal static electric field sections displayed in red. The other details from the left-panel figure 
are omitted in the right-panel figure. Both figures are from Ref.~\cite{Suleiman:2023whz}.}
\label{fig:fig_specific_edm}
\end{figure}

For the eEDM measurement, the simplest two-energy ST ring's schematic conforming to the above principles is shown in the left panel of Fig.~\ref{fig:fig_specific_edm}. It is also quite compact, partly because 
of being a simple setup. Compactness is important for maximizing the rate of the eEDM signal accumulation. The left-panel ring consists of two low-energy and two high-energy arcs. The beam directions in the two 
arcs of each energy are opposite making the net bending angle zero. The low- and high-energy arcs are connected to potentially common capacitor plates providing acceleration and deceleration of 600~kV. 
A bunching RF cavity is located in a low-energy section to minimize its voltage requirements. The entire ring consists of arc sections, where the electric field remains transverse to the reference electron 
velocity, and of straight sections in which the electric field is longitudinal to the beam direction. We consider the spin dynamics in these two cases separately. The bunching RF fields do not contribute to 
this scenario because their average first-order spin effect is zero. 
\FloatBarrier

In spite of these promising outcomes, an eEDM ring optics design must also provide long spin coherence time, adequate momentum acceptance \& dynamic aperture, low beam emittance growth rates due to intra-beam
scattering (IBS), and reasonable stored beam size. All of this could be not so efficient to achieve just by using the conceptual design. Consequently, we choose the so-called Bates design~\cite{Flanz:1985lnw} 
for the eEDM ring arcs, where each arc consists of five alternating bends located symmetrically about its center, such as 
$\pm | \theta_1 | \wedge \mp | \theta_2 | \wedge  \pm 2| \theta_3 | \wedge  \mp | \theta_2 | \wedge \pm | \theta_1 |$ with $\theta_i$ being the three different angular parameters defining the arc's geometry. 
We use half-circular arcs such that, for each arc, $\left | 2\sum_{i = 1}^3 \theta_i \right | = 180^\circ$.

Each arc can be made very compact with most of its length occupied by bending elements, which is important for maximizing the eEDM spin rotation rate. 
The Bates optics structure of the arcs of the ST ring is based on the weak-focusing principle that provides robustness and allows for large beam size and momentum spread. So, the horizontal focusing comes 
solely from the arc's bending curvature while the vertical focusing is set by a weak vertical-focusing gradient of the electric field. Such a design is relatively simple from the engineering point of view 
by having only a few adjustment knobs. Besides, by not having large beam envelope variations, the Bates design is also efficient in terms of generating minimal IBS growth rates. The IBS is the main mechanism 
for degradation of intense electron beams at considered low energies. In addition, we note that the dispersion is controlled by changing the bending direction. 

We assume longitudinal fields of $5$~MV/m and bending fields of $\leq 10$~MV/m on the design orbit. Small beam sizes are required to have sufficient beam aperture and to maintain long beam lifetime.  
The corresponding footprint of the specific ST ring is shown in the right panel of Fig.~\ref{fig:fig_specific_edm}. Some of the ring design and beam parameters are summarized in Table~\ref{tab:ring-param}.
\begin{table*}[!htb] \caption{Several parameters of the specific ST ring under consideration, supplemented by the IBS growth times and maximum rms beam size, both shown when 
the stochastic cooling is turned on. See Ref.~\cite{Suleiman:2021whz} for many more details on the entire ring optics and on other ring and beam parameters.}
\label{tab:ring-param}
\centering
\begin{tabular}{lll}
\hline \hline
Quantity                 & \multicolumn{2}{l}{Value} \\ \hline
Ring circumference	     & \multicolumn{2}{l}{3.55~m} \\ 
Circulation frequency $f_c$	 & \multicolumn{2}{l}{68.5~MHz} \\
Straight section length	 & \multicolumn{2}{l}{12.3~cm} \\ 
Beam pipe aperture $\pm a_x$  & \multicolumn{2}{l}{$\pm 3$~cm} \\ 
IBS growth times $\tau_{x}^{\rm IBS}$/$\tau_{y}^{\rm IBS}$/$\tau_{z}^{\rm IBS}$ & \multicolumn{2}{l}{40\,s/40\,s/4\,s} \\
\hline \hline
Section                      & Low-energy                 & High-energy \\ \hline
Lorentz energy factor $\gamma$    & 1.4 & 2.6 \\
Arc bending radii $\rho_i$ & $\pm 8.3$, $\mp 10.8$, $\pm 9.2$~cm & $\pm 11.3$, $\mp 11.3$, $\pm 11.3$~cm \\
Maximum rms beam size $\sigma_{x}^{\rm max}$/$\sigma_{y}^{\rm max}$ & 5.9/5.6~mm & 3.0/5.6~mm \\
\hline \hline
\end{tabular}
\end{table*}
\FloatBarrier

A straightforward way to obtain the eEDM spin rotation per turn $N$ around the ring, $\frac{\partial \left| \psi_{\rm EDM} \right|}{\partial N}$, is to treat the eEDM signal as a perturbation of the MDM 
spin motion  on the closed orbit~\cite{Filatov:2020ygb} (as mentioned at the beginning of this section). The magnitude of the spin rotation due to the eEDM in one turn can be obtained as
\begin{equation}
\frac{\partial \left| \psi_{\rm EDM} \right|}{\partial N} = \left|2 \eta \left[ \frac{\gamma_2^2 \beta_2}{1-\gamma_2^2 \beta_2^2 G} - \frac{\gamma_1^2 \beta_1}{1-\gamma_1^2 \beta_1^2 G} - \ln \frac{\gamma_2 + \sqrt{\gamma_2^2 - 1}}{\gamma_1 + \sqrt{\gamma_1^2 - 1}} \right] \sin\left( \frac{\omega_M^1}{2} \pi\right) \sin\left( \frac{\omega_M^2}{2} \pi\right) \right| ,
\label{eq:eq_R}
\end{equation}
where the quantity $\omega_M^n$ stands for
\begin{equation} 
\omega_M^n = - \frac{1+G-\gamma_n^2 G}{\gamma_n} ,
\label{eq:eq_Q}
\end{equation}
and where $n=1$, 2 specifies the electron Lorentz energy factor $\gamma_n$ in the two different energy regions (with the spin rotation axis lying on the horizontal plane). The physical meaning 
of $\omega_M^n$ is the spin precession due to the MDM per unit orbital angle in transverse electric field.  A detailed derivation of Eq.~(\ref{eq:eq_R}) can be found in Ref.~\cite{Suleiman:2021whz}.

We also assume that the accelerating/decelerating electric fields constitute only a half of the ring circumference, i.e., a packing factor of 0.5. The rest of the ring is occupied by other machine 
components and experimental apparatus. At this point, it is relevant to indicate that the electron revolution time in this two-energy ST ring (either conceptual or specific) is given by
\begin{equation}
T= 2 \frac{mc^2/e}{c |E|} \left( 2 \pi \gamma_1 \beta_1 + 2 \pi \gamma_2 \beta_2 + 4 \left[\sqrt{\gamma_2^2-1} - \sqrt{\gamma_1^2-1} \right] \right) .
\label{eq:eq_T}
\end{equation}
In turn, the EDM spin rotation per unit time is the following:
\begin{equation}
\frac{\partial^2 \left| \psi_{\rm EDM} \right|}{\partial \eta \partial t} = \frac{1}{T} \frac{\partial^2 \left| \psi_{\rm EDM} \right|}{\partial \eta \partial N} .
\label{eq:eq_U}
\end{equation}

The difference corresponding to $\gamma_{1} = 1.4$ and $\gamma_{2} = 2.6$, i.e., the electrostatic potential difference over the accelerating/decelerating gaps corresponds to 600~kV, as shown in 
the left panel of Fig.~\ref{fig:fig_specific_edm}. In Table~\ref{tab:ring-comp}, we compare the results obtained from the conceptual/specific ST ring approaches with those obtained from the so-called 
``magic energy" (ME) approach \cite{Farley:2003wt}. The ME approach is based on an all-electric ring with $B_{y} = 0$ and $\gamma^{2} = 1 + 1/G$ (see Eq.~(\ref{eq:T-BMT_labFrame-ME})), and it works 
only for $G > 0$ (e.g., proton or electron) at a very specific energy\footnote{With the ME approach, two experiments have been proposed for pEDM measurements with a sensitivity of $10^{-29}~e$\,cm 
in rings at 232.8~MeV with $\geq 500$~m circumference \cite{Anastassopoulos:2015ura,Haciomeroglu:2018nre,1865072,Omarov:2020kws,CPEDM:2019nwp}.}.
\setlength{\tabcolsep}{10pt}
\begin{table*}[!htb] \caption{Comparison of the EDM spin rotations in the ME and two-energy ST rings.}
 \label{tab:ring-comp}
    \centering
    \begin{tabular}{l|c|c|c}
\hline \hline
Scheme           & Energy  & $\partial^2 \left| \psi_{\rm EDM} \right|/(\partial \eta \partial N)$ &  $\partial^2 \left| \psi_{\rm EDM} \right|/(\partial \eta \partial t)$ \\ 
                 &  configuration $(\gamma)$  & (rad)        & ($\times 10^9$~rad/s) \\ 
 \hline 
 ME & 29.38            & 92.24        & 1.47 \\ 
 ST (conceptual) & (1.4, 2.6)  & 4.24         & 0.46 \\ 
 ST (specific) & (1.4, 2.6)  & 4.24         & 0.29 \\ 
 \hline \hline
    \end{tabular}
\end{table*}

However, when we compare the ME ring results with the conceptual \& specific ST rings' results in Table~\ref{tab:ring-comp}, it is only a hypothetical exercise. The reason is that there are 
currently no proposals for eEDM measurements in a storage ring at a magic energy. The primary difficulties with applying the magic-energy approach to electrons are lack of suitable polarimetry 
and complications arising because of synchrotron radiation. Although, the \textbf{specific design} is the winner even in such a comparison.

\bigskip
\section{Electron EDM measurement precision}
\label{sec:prec_edm}

We refer to Ref.\cite{Suleiman:2021whz} for detailed discussions on the beam physics limitations, spin stability and precision control, as well as on the polarized electron source and Mott polarimetry
to be used in the eEDM measurement under consideration. Besides, see also the same reference for details on one of the very possible schemes for operating the eEDM ST \textbf{specific} storage ring and 
performing Mott polarization measurements.

\paragraph{Expected statistical limit.}
The eEDM will accumulate a positive vertical polarization component for filled two beam bunches (${\it CRA}$,+h) and (${\it CRB}$,-h), as well as a negative vertical component for the other filled two bunches 
(${\it CRA}$,-h) and (${\it CRA}$,+h), where $\pm$h is the positive/negative helicity.  After total running of 1825 fills (or five years), one can achieve a measurement of vertical polarization due to the 
eEDM spin precession to statistical uncertainty of $\delta P \approx 4.7 \times 10^{-6}$~rad or \SI{4.7}{\micro\radian}. For the given polarization measurement scheme, the statistical uncertainty of the eEDM measurement 
per fill can be calculated as~\cite{CPEDM:2019nwp,Kim:2021pld}:
\begin{equation}
\sigma_{\rm EDM} = \sqrt{8}\,\frac{d_e}{\sqrt{Q/e \; \epsilon \;} \; A_y \; P \; \left(\partial^2 \left| \psi_{\rm EDM} \right|/(\partial \eta \partial t)\right) \; {\rm SCT} } . 
\label{eq:EDM_limit}  
\end{equation}
With the numerical values given in Table~\ref{tab:pol_par}, the statistical uncertainty per fill on the EDM measurement is about $2.5 \times 10^{-28}~e$\,cm. The spin precession rate 
$\partial^2 \left| \psi_{\rm EDM} \right|/(\partial \eta \partial t)$ is shown in Table~\ref{tab:ring-comp}.
\begin{table*}[!htb] \caption{Parameters for evaluating the statistical uncertainty of the eEDM measurement.}
 \label{tab:pol_par}
    \centering
    \begin{tabular}{lcl}
\hline \hline
Electrons per fill                & $N_e=Q/e$         & $5.0 \times 10^{10} \,(2.5 \times 10^{10}$ {\it CRA}, \,$2.5 \times 10^{10}$ {\it CRB}) \\
Polarimeter efficiency            & $\epsilon$    & 0.0024 \\
Analyzing power 	              & $A_y$         & 0.45 \\
Longitudinal polarization         & $P$           & 0.90 \\
Spin coherence time               & SCT           & 1~day (86400~s) \\
\hline
Beam lifetime                     & BL            & $\gtrsim$ 1~day \\
\hline \hline
   \end{tabular}
\end{table*}
The respective projected statistical limit after five years of data taking per one eEDM ST ring is about $5.8 \times 10^{-30}~e$\,cm with 90\% C.L.. But the expectation is that further optimization and 
improvements will still decrease this limit. How this can be accomplished will be highlighted and described thoroughly in a future proposal.  

\paragraph{Systematic uncertainty.}

The MDM spin rotation limits the tiny eEDM signal that can be measured and introduces systematic uncertainties \cite{Omarov:2020kws,Carli:2022ftv}. In particular, the former should average to zero over 
a single turn in the eEDM ST ring, nevertheless, fringe and environmental electromagnetic fields plus errors in the construction and alignment of the ring elements may introduce non-zero MDM spin rotations. 
But there are good approaches to further suppress these residual effects.
\begin{itemize}
\item[(i)] First approach relies on the state-of-the-art shielding of the environmental fields, where the small size of the storage ring makes shielding very practical\footnote{For the discussed low-energy 
electron beams combined with the proposed vacuum chamber design, some shielding effect is expected to occur, which will suppress the emission of synchrotron radiation \cite{Murphy:1996yt,Warnock:1988sb}, 
such that the bunching cavity can easily compensate for this energy loss.};

\item[(ii)] Second approach relies on the fact that the EDM effect is time-reversal violating, whereas the majority of the experimental systematic effects are time-reversal conserving. The time-reversal 
of the measurement technique can be implemented by the use of two $CR$ beams.

\item[(iii)] Third approach is based on suppression of systematic uncertainties by reversing the direction of the electron spin and by combining data, e.g., collected during eEDM measurements, such as 
$\frac{ {\rm ({\it CRA},+h)} - {\rm ({\it CRA},-h)} + {\rm ({\it CRB},-h)} - {\rm ({\it CRB},+h)}}{4}$.

\item[(iv)] Fourth approach will use two of the four bunches (with either radial or vertical polarization) in each direction of the ring to control background electromagnetic fields.
\end{itemize}

Not all systematic effects cancel with either spin reversals or beam velocity because some of those effects mimic the EDM signal, and it will require more study and scrutiny~\cite{Omarov:2020kws,CPEDM:2019nwp}.
For example, a radial background magnetic field that generates a vertical spin rotation identical to the eEDM. It is also relevant to mention that the expected systematic uncertainty stemming from the Mott 
polarimetry is $\delta P \approx~$\SI{1}{\micro\radian}. This is similar to the expected value from proton polarimetry in the COSY experiment~\cite{Brantjes:2012zz}.

Our colleagues have invested considerable efforts to study different kind of systematics for the pEDM measurements \cite{Anastassopoulos:2015ura,Omarov:2020kws,CPEDM:2019nwp}. The proposed eEDM measurements 
will also benefit from those studies, and we assume that eventually the final eEDM projection will be mostly statistically limited. The systematic uncertainty budget will be presented in the future proposal, nevertheless, there may still be some non-suppressible systematic uncertainties, which can be studied at the time of the actual eEDM experiment.

\bigskip
\section{Search for axions with a spin-transparent storage ring using low-energy electron beams}
\label{sec:search_axion}

In Sec.~\ref{sec:axion} we already provided some introduction to axions, and here we wish to concisely describe how these ultralight dark matter candidates can be searched using a similar type 
of setup discussed in Sec.~\ref{sec:edm_ring}. We also note that originally it is an idea coming from the ME-based design of storage rings~\cite{Graham:2020kai} to look for spin precession 
induced by dark matter and dark energy. 

In general, the field gradient of low-mass axions couples to the spin of transversely-polarized electrons stored in a ring with a sensitivity proportional to the relativistic beam velocity 
and the beam Spin Coherence Time. As such, low-energy spin-transparent storage rings can be used to measure spin precession frequency induced by axion fields. In an eEDM ring, one could use 
longitudinally polarized electrons to measure the eEDM, whereas transversely polarized electrons would be used in an axion ring to search for axions. Thereby, the axion ring is analogous to the eEDM 
ring, and it is a Figure-8 ring without longitudinal electric fields. In such axion search experiments, the spin rotates around the electron’s velocity and the main systematic uncertainty will be 
caused by the longitudinal background magnetic field that rotates the spin of the counter-rotating $A$ and $B$ beams in the same direction. However, the axion interaction will rotate the spin in 
opposite directions, resulting in cancellation of this background effect when combining data from both beams.

One can specifically consider a sensitivity of a ST ring to axion gradients sourced by the earth or test masses through the monopole interaction with the earth nucleons. The dipole interaction with 
electrons appears as an effective background magnetic field in an axion gradient background, which generates spin precession. The force that affects the electron motion from the axion interaction 
is extremely small. But a Figure-8 axion ST storage ring can be used to search for spin precession induced by the axion interaction with a measurable rate down to
$0.2 \cdot 10^{-9}$~rad/s (or 0.2~nrad/s) 
per one ring.

\begin{figure}[h!]
\centering
\hspace{0.0cm}
\includegraphics[width=11.5cm]{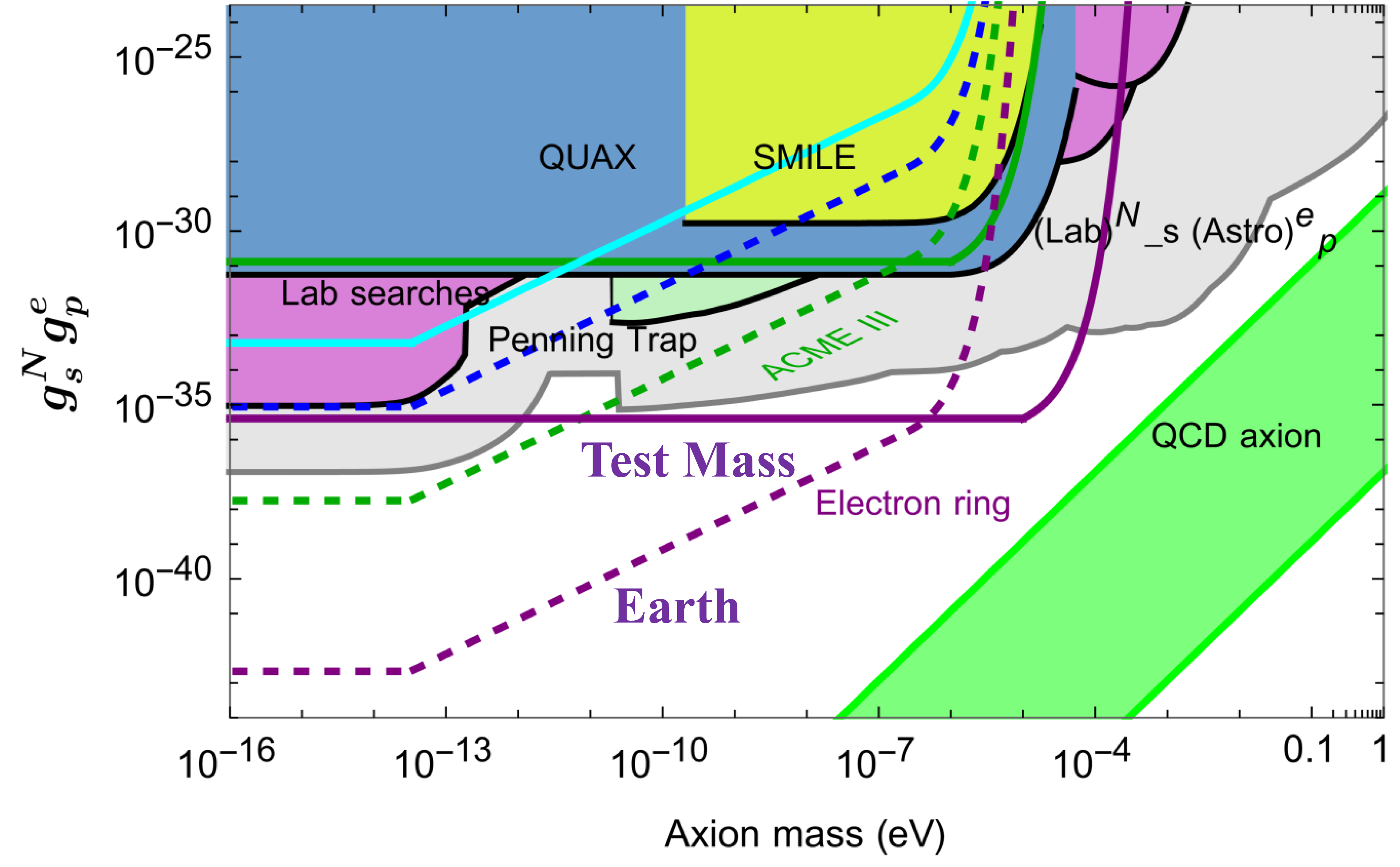}
\caption{Axion-mediated monopole-dipole forces exerted on electrons at various spin precession experiments with their bounds projected on the parameter space of scalar-pseudoscalar nucleon-electron 
couplings vs. axion mass. If the magenta solid and dashed lines are combined, it provides the strongest sensitivity (obtained with one axion ring) compared to other experiments. This figure is from
Ref.~\cite{Suleiman:2024hvh}.}
\label{fig:fig_axion_plot}
\end{figure}

Figure~\ref{fig:fig_axion_plot} demonstrates the sensitivity to axion-mediated forces in the axion ST ring sourced from the earth or from test masses in the lab. In particular, the QCD axion band 
is shown in green. The cyan line corresponds to the current sensitivity coming from Penning trap experiments \cite{Fan:2023hci}. The dashed blue and dashed green lines correspond to future improvements. 
The magenta solid line corresponds to a sensitivity stemming from cubic lead bricks of 10~cm size arranged around an axion ring at a distance of 10~cm from the beam, covering only 10\% of the ring 
circumference (the sensitivity increases linearly with this fraction). Then the magenta dashed line corresponds to the limits obtained with the axion field from the earth nucleons. The combination 
of both magenta configurations gives the strongest bounds, which surpass any existing and foreseeable search by several orders of magnitude. More details can be found in 
Refs.~\cite{Fan:2023hci,Agrawal:2022wjm,Agrawal:2023lmw}.

\bigskip
\section{Spin-transparent storage ring concept's development for low-energy polarized positron beams at JLab}
\label{sec:pos_edm}

In reality, the application of the discussed ST storage rings to the positron EDM measurements would be quite similar to the case of the eEDM measurement. 
For instance, producing and accumulating polarized positron bunches are achievable in an $e^{+}$EDM ring, with the same systematic effect suppression procedure to be applied. 
Nonetheless, there are also important differences to consider (mostly not related to the ring's structure).
\begin{enumerate}
\item[(a)] The positron Mott polarimeter uses exactly the same physical mechanism but for positrons instead of electrons; thus, to measure the positrons' transverse spin polarization.
The asymmetry formula is the same, $A = S(\theta, E) P$, however, the sign and magnitude of the Sherman function $S(\theta, E)$ (analyzing power) change somewhat because
of the opposite charge of the positron. 

\item[(b)] The key physics difference is that electrons and positrons interact differently with the nuclear Coulomb field: the electrons are attracted to a nucleus, whereas 
the positrons are repelled from the nucleus. This modifies (i) the scattering phase shifts, (ii) the spin-orbit interaction strength, (iii) the Sherman function value. As a result, 
the positron Sherman function \cite{Jakubassa:2026} is usually slightly smaller, multiple scattering corrections differ, and $e^{+}$ and $e^{-}$ cross sections also differ from 
each other with lower statistical precision for the positrons.  

\item[(c)] When the positrons hit a target, 511~keV gamma rays are produced from the annihilation reaction $e^{+} + e^{-} \rightarrow 2\gamma$, which can create
larger backgrounds. The electron Mott polarimetry does not have such backgrounds.

\item[(d)] Since a polarized positron beam requires a multi-step process for its production, the produced beam may ultimately have lower polarization and larger energy spread 
compared to the a polarized electron beam.  

\item[(e)] JLab has an electron Mott polarimeter at its disposal and operates multiple electron injector systems to support its diverse research programs. Meanwhile, JLab 
currently does not have a positron injector; accordingly, the positron polarimetry is not developed yet.  
\end{enumerate}

\medskip
After a positron injector becomes available at JLab, it will trigger the positron polarimetry development. Notwithstanding the absence of the positron Mott polarimeter currently, if it 
were to be available at some point later and if we used the characteristics as well as the beam parameters similar-to those of the eEDM ring (but for positrons), the precision of an 
$e^{+}$EDM measurement over 5-years of data taking with only one positron ST storage ring would be \,$\sim 5 \times 10^{-29}~e$\,cm. 

Relevant to and as a continuation of our discussion in Sec.~\ref{sec:motiv}, let us emphasize that a far-reaching goal would be a comparison of the measured EDMs of the electron and positron 
in the future. Such a comparison could
\begin{enumerate}
\item[(i)] provide an important independent experimental cross-check: different storage-ring systematics could affect the electron and positron measurements dissimilarly because of the 
opposite charge of these two particles;

\item[(ii)] test the $CPT$ symmetry with different EDM observables: $CPT$ has been tested extremely precisely in masses and magnetic moments, however, much less so for EDMs;

\item[(iii)] produce different EDM contributions for electrons and positrons coming from underlying mechanisms based on some new-physics models (for instance, involving the Lorentz 
violation \cite{Aghababaei:2024hcb,Bluhm:1998wf} or $CPT$-violating background fields \cite{Ding:2026aac}): the comparison would strongly constrain such models.
\end{enumerate}

Thereby, the primary motivation here is to perform one of the most stringent $CPT$ tests possible in the lepton sector. That is why a future low-energy ST-storage-ring based experiment for measuring
the $e^{+}$EDM will have a distinct justification beyond simply repeating the eEDM measurement.

\bigskip
\section{Summary}
\label{sec:con}
In this paper we presented a beyond-state-of-the-art design of distinctly specialized Figure-8 table-size ($\sim$ 1\,m) all-electric spin-transparent storage rings, which use 
low-energy polarized electron beams and Mott polarimetry. The design based on the transparent spin methodology cancels the spin precession induced by the magnetic dipole moment (at any 
beam energy), whereas allows for the spin precession due to the fundamental physics of interest to accumulate. 

The key features of our new approach discussed here and also covered in great details in Refs.~\cite{Suleiman:2023whz,Suleiman:2021whz} can be summarized 
as follows:
\begin{enumerate}
\item[A1] The electron permanent eEDM measurement can be carried out in low-energy and two-energy ST storage rings;
\item[A2] The use of Mott polarimetry will be efficient for building up the vertical component of electron beam polarization in the considered beam energy range of $~\lesssim1~{\rm MeV}$;
\item[A3] The projected statistical limit for the eEDM is already shown in the Executive Summary, Sec.~\ref{sec:motiv}, and Sec.~\ref{sec:prec_edm};
\item[A4] A longitudinally polarized electron ring (the eEDM ring) can be used for measuring the permanent eEDM while a transversely polarized electron ring (axion ring) can be used for
searching for axions;
\item[A5] The axion ring can measure spin precession frequency induced by the axion field with a sensitivity bound shown in the Executive Summary, Sec.~\ref{sec:motiv}, and Sec.~\ref{sec:search_axion}.
\end{enumerate}

\medskip
\hskip -0.55truecm
The proposed experimental setup has the following additional characteristics too: 
\begin{itemize}
\item[B1] The ST configuration of a Figure-8 storage ring makes the spin tune (precession) energy-independent and insensitive to beam emittances;
\item[B2] The small ring orbit size allows for efficient stochastic cooling, and the ring imperfections including background magnetic fields are well suppressed;
\item[B3] The Spin Echo technique \cite{Derbenev:2021} can be applied to compensate spin decoherency caused by the  beam emittances, and may further advance the precision of spin control;
\item[B4] These rings provide long Spin Coherence Time ($\sim 1$~day) and operate at any energy including low energies;
\item[B5] There are two counter-rotating beams (to be bunched and unbunched) with various spin and polarization orientations providing good control of systematic uncertainties;
\item[B6] Synchrotron radiation is expected to be suppressed from shielding effects coming from the combination of the electron beam's low energy with the proposed vacuum chamber 
design; and there will be minimal safety issues (no radiation and activation); 
\item[B7] The ring has a small easily-manageable size, much lower cost (compared to pEDM rings' cost), and can be potentially used as a powerful testbed for larger-scale similar kind of experiments.
\end{itemize}

\section*{Acknowledgements}
\label{sec:ack}

This material is based upon work supported by the U.S. Department of Energy, Office of Science, Office of Nuclear Physics under Contract No. 89243126CSC000213, as well as supported by the startup 
funds awarded to Prof. Wenliang (Bill) Li by the Department of Physics and Astronomy at Mississippi State University.

\bigskip

\end{document}

%% file: title.tex
\begin{titlepage}

\begin{flushright}
JLAB-ACC-26-5022
\end{flushright}

\vspace{0.25in}
\begin{center}
\textup{
{\LARGE {\bf High Precision Fundamental Physics Experiments at JLab with Spin-transparent Storage Rings of Low-energy Polarized Electron Beams}
}
}
\end{center}

\begin{center}
\vspace{0.65in}

{\bf 
Riad Suleiman\affmark[a] \orcidlink{0000-0001-6767-1579},  
Silviu Covrig Dusa \orcidlink{0000-0001-9117-8493},  
Yaroslav Derbenev,  
\\
Joseph Grames \orcidlink{0000-0001-9390-8752},  
Tyler J. Hague \orcidlink{0000-0003-1288-4045},  
Ching Him Leung \orcidlink{0000-0001-7907-3728},  
Michael Nycz \orcidlink{0000-0002-3570-9103}  
\\
}
\smallskip
{\it Thomas Jefferson National Accelerator Facility, Newport News, Virginia 23606, USA}
\vspace{0.35in}

{\bf 
Vladimir Khachatryan\affmark[b] \orcidlink{0000-0003-2567-2930},  
Wenliang (Bill) Li \orcidlink{0000-0002-8108-8045},  
Devesh Bhattarai \orcidlink{0009-0007-7847-2533},  
\\
Preet Mann,  
Zakia Tithi \orcidlink{0009-0006-4056-1402},  
Erik Wrightson \orcidlink{0009-0004-4889-1751},  
Zichen Yin \orcidlink{0009-0001-3710-3033}  
\\
}
\smallskip
{\it Mississippi State University, Starkville, Mississippi 39762, USA}
\vspace{0.35in}

{\bf 
Kurt Aulenbacher \orcidlink{0000-0001-5433-7424}  
\\
}
\smallskip
{\it Helmholtz-Institute Mainz, 55128 Mainz, Germany} \\
\smallskip
{\it GSI Helmholtzzentrum für Schwerionenforschung GmbH, 64291 Darmstadt, Germany}
\vspace{0.35in}

{\bf 
Mario Reig \orcidlink{0000-0001-6815-5605}  
\\
}
\smallskip
{\it  CERN, CH-1211 Geneva, Switzerland}
\vspace{0.35in}

{\bf 
Alberto Accardi \orcidlink{0000-0002-2077-6557}  
\\
}
\smallskip
{\it Christopher Newport University, Newport News, VA 23606, USA} \\
\smallskip
{\it Thomas Jefferson National Accelerator Facility, Newport News, Virginia 23606, USA}
\vspace{0.35in}

{\bf 
Howard Oh \orcidlink{0000-0001-7372-6695}  
\\
}
\smallskip
{\it Defense Physics Laboratory, Princeton, New Jersey, NJ 08542, USA}
\vspace{0.35in}

{\bf 
Forrest Friesen \orcidlink{0000-0002-1854-6306},  
Ye Tian \orcidlink{0000-0002-9979-0641}  
\\
}
\smallskip
{\it Duke University, Durham, North Carolina 27708, USA}
\vspace{0.35in}

{\bf 
Paul Guèye \orcidlink{0000-0002-0199-9501}  
\\
}
\smallskip
{\it Facility for Rare Isotope Beams, Michigan State University, \\
East Lansing, Michigan 48824, USA} \\
\smallskip
{\it Michigan State University, East Lansing, Michigan 48824, USA}
\vspace{0.35in}

{\bf 
Marco Schreck \orcidlink{0000-0001-6585-4144}  
\\
}
\smallskip
{\it Universidade Federal do Maranhão, São Luís, Maranhão 65080-805, Brazil}
\vspace{0.35in}

{\bf 
Magno Valério Trindade Machado \orcidlink{0000-0003-2821-8266}  
\\
}
\smallskip
{\it Universidade Federal do Rio Grande do Sul, Porto Alegre, \\
Rio Grande do Sul 90010-150, Brazil}
\vspace{0.35in}

{\bf 
Alexandre Vieira \orcidlink{0000-0002-5583-1560}  
\\
}
\smallskip
{\it Universidade Federal do Triângulo Mineiro, Iturama, Minas Gerais 38280-000, Brazil}
\vspace{0.35in}

{\bf 
Udit Raha \orcidlink{0000-0003-0234-8767}  
\\
}
\smallskip 
{\it Indian Institute of Technology Guwahati, Guwahati, Assam 781039, India}
\vspace{0.35in}

{\bf 
Marco Battaglieri \orcidlink{0000-0001-5002-8771}  
}
\\
\smallskip
{\it Istituto Nazionale Fisica Nucleare - Sezione di Genova, 16146 Genova, Italy} \\
\smallskip
{\it Università degli Studi di Genova, 16126 Genova, Italy}
\vspace{0.35in}

{\bf 
Nathaniel Sherrill \orcidlink{0000-0001-6909-3872}  
\\
}
\smallskip
{\it Institut für Theoretische Physik, Leibniz Universität, Hannover, 30167, Germany}
\vspace{0.35in}

{\bf 
Vasiliy Morozov \orcidlink{0000-0003-2352-8759}  
\\
}
\smallskip
{\it Oak Ridge National Laboratory, Oak Ridge, Tennessee 37831, USA}
\vspace{0.35in}

{\bf 
Carlos Ayerbe Gayoso \orcidlink{0000-0001-8640-5380}  
\\
}
\smallskip
{\it Old Dominion University, Norfolk, Virginia 23529, USA}
\vspace{0.35in}

{\bf 
Eric Fuchey \orcidlink{0000-0003-0100-6052}  
\\
}
\smallskip
{\it Physics Department, College of William and Mary, Williamsburg, Virginia 23187, USA}
\vspace{0.35in}

{\bf 
Son Nguyen \orcidlink{0000-0002-6104-7035}  
\\
}
\smallskip
{\it Washington and Lee University, Lexington, Virginia 24450, USA}
\vspace{0.35in}

\end{center}

\begin{center}
\affaddr{a -- suleiman@jlab.org, ~~b -- vk355@msstate.edu/vladkha@jlab.org} \\
\end{center}

\newpage
\section*{Executive Summary}

This paper aims to present specialized high-precision fundamental physics experiments based on applying compact particle accelerator spin-transparent 
storage rings.

\medskip
\hskip -0.6truecm
The main physics goals are to directly measure the electron's permanent electric dipole moment (eEDM) and search for ultralight dark matter candidates (such as axions), 
using data accumulated in small-sized ($\sim$ 1\,m) storage rings. 

\medskip
\hskip -0.55truecm
The measured observables will be 
\begin{itemize}
\item[(i)] Permanent eEDM ($d_{e}$) extracted from measuring the electron's beam polarization and then the eEDM spin precession rate. The respective projected statistical 
limit after five years of data taking per one eEDM spin-transparent ring is about $5.8 \times 10^{-30}~e$\,cm with 90\% C.L.. This number will be close to the hitherto-best-measured 
upper bound $d_{e} < 4.1 \times 10^{-30}~e$\,cm with 90\% C.L., extracted indirectly from a measurement using HfF$^{+}$ ions. Albeit at a lower accuracy, one-ring direct measurement 
could serve as an independent sanity check for upcoming eEDM indirect measurements.

Meanwhile, using multiple rings would only improve (or much improve) the precision of the eEDM direct measurements, depending on the exact number of usable rings and data-taking 
period. As such, the one-ring eEDM experiment will additionally show a Proof of Concept, providing an excellent testbed for the purpose of fully testing and adopting this technology, 
then moving on to employing multiple-ring platforms in the future.

\item[(ii)] Spin precession rate induced by the spinless axion field measured through the axion field gradient coupled to transversely polarized electron spin, which will 
then be mapped onto the parameter space of scalar-pseudoscalar nucleon-electron couplings versus axion mass to be compared with the results from other worldwide experiments.
Just one-ring-measured precession rate 
induced by the axion interaction will be as low as $0.2 \cdot 10^{-9}$~Rad/s (or, 0.2~nHz), and the sensitivity obtained by the axion spin-transparent storage ring will exceed any 
existing and anticipated search by several orders of magnitude.
\end{itemize}

\medskip
\hskip -0.55truecm
The experimental platform is suggested to be hosted at the Low Energy Recirculator Facility (LERF) of JLab. The existing electron injector at the LERF can be used for both experiments.
\begin{itemize}
\item[$\bullet$] Cost of one storage ring for the electron EDM measurement: \$7.5M; 
\item[$\bullet$] Cost of one storage ring for the axion search: \$7.5M;
\item[$\bullet$] Total cost: \$15M.
\end{itemize}

\smallskip
\hskip -0.6truecm
Furthermore, the eEDM ring platform can also be used to directly measure the positron's permanent EDM, by having a far-reaching goal to compare the EDMs of the electron and positron in the future. 

\medskip
We hope that our proposed small-scale but largely competitive experiments will open a new avenue to further strengthen JLab's goal in becoming a multipurpose research facility, 
and in order to lead global intensity-frontier physics \& cutting-edge technological innovation programs. As such, we are seeking feedback and endorsement from the PAC
to promote the experiments elaborated in this paper and to start working on a follow-up proposal.

The experiments we propose will also become a very good training ground for students, postdocs and juniors specializing in accelerator physics and technology. 

\end{titlepage}

%% file: EDM_LOI.bbl
\begin{thebibliography} {99}

\bibitem{Suleiman:2023whz}
R.~Suleiman, V.~S.~Morozov and Y.~S.~Derbenev,
``High precision fundamental physics experiments using compact spin-transparent storage rings of low energy polarized electron beams,''
Phys. Lett. B \textbf{843}, 138058 (2023).

\bibitem{Suleiman:2021whz}
R.~Suleiman, V.~S.~Morozov and Y.~S.~Derbenev,
``On Possibilities of High Precision Fundamental Physics Experiments in Spin-Transparent Storage Rings of Low Energy Polarized Electron Beams,''
[arXiv:2105.11575 [physics.acc-ph]].

\bibitem{Suleiman:2024hvh}
R.~Suleiman, Y.~Derbenev, M.~Grau and V.~Morozov,
``Particle accelerator spin-transparent storage rings for beyond state-of-the-art science,''
JACoW \textbf{IPAC2024}, WEAN2 (2024),
\url{https://digitalcommons.odu.edu/cgi/viewcontent.cgi?article=1901&context=physics_fac_pubs}.

\bibitem{Suleiman:2026zxn}
R.~Suleiman, V.~Morozov and M.~Grau,
``Spin-transparent storage rings for quantum computing,''
JACoW \textbf{NAPAC2025}, THP076 (2026),
\url{https://inspirehep.net/files/bf2c804e76de7725814af8ee9d5cd6e5}.

\bibitem{Khriplovich:1997ga}
I.~B.~Khriplovich and S.~K.~Lamoreaux,
``CP violation without strangeness: Electric dipole moments of particles, atoms, and molecules,''
Texts and Monographs in Physics, Springer, Berlin, Heidelberg (1997).

\bibitem{Yamaguchi:2020eub}
Y.~Yamaguchi and N.~Yamanaka,
``Large long-distance contributions to the electric dipole moments of charged leptons in the standard model,''
Phys. Rev. Lett. \textbf{125}, 241802 (2020)
[arXiv:2003.08195 [hep-ph]].

\bibitem{Ema:2022yra}
Y.~Ema, T.~Gao and M.~Pospelov,
``Standard Model Prediction for Paramagnetic Electric Dipole Moments,''
Phys. Rev. Lett. \textbf{129}, no.23, 231801 (2022)
[arXiv:2202.10524 [hep-ph]].

\bibitem{Dine:2003ax}
M.~Dine and A.~Kusenko,
``The Origin of the matter - antimatter asymmetry,''
Rev. Mod. Phys. \textbf{76}, 1 (2003)
[arXiv:hep-ph/0303065 [hep-ph]].

\bibitem{Sakharov:1967dj}
A.~D.~Sakharov,
``Violation of CP Invariance, C asymmetry, and baryon asymmetry of the universe,''
Pisma Zh. Eksp. Teor. Fiz. \textbf{5}, 32-35 (1967).

\bibitem{Gavela:1993ts}
M.~B.~Gavela, P.~Hernandez, J.~Orloff and O.~Pene,
``Standard model CP violation and baryon asymmetry,''
Mod. Phys. Lett. A \textbf{9}, 795-810 (1994)
[arXiv:hep-ph/9312215 [hep-ph]].

\bibitem{Engel:2013lsa}
J.~Engel, M.~J.~Ramsey-Musolf and U.~van Kolck,
``Electric Dipole Moments of Nucleons, Nuclei, and Atoms: The Standard Model and Beyond,''
Prog. Part. Nucl. Phys. \textbf{71}, 21-74 (2013)
[arXiv:1303.2371 [nucl-th]].

\bibitem{Chupp:2017rkp}
T.~Chupp, P.~Fierlinger, M.~Ramsey-Musolf and J.~Singh,
``Electric dipole moments of atoms, molecules, nuclei, and particles,''
Rev. Mod. Phys. \textbf{91}, no.1, 015001 (2019)
[arXiv:1710.02504 [physics.atom-ph]].

\bibitem{Yamanaka:2016umw}
N.~Yamanaka,
``Review of the electric dipole moment of light nuclei,''
Int. J. Mod. Phys. E \textbf{26}, no.4, 1730002 (2017)
[arXiv:1609.04759 [nucl-th]].

\bibitem{Pospelov:2005pr}
M.~Pospelov and A.~Ritz,
``Electric dipole moments as probes of new physics,''
Annals Phys. \textbf{318}, 119-169 (2005)
[arXiv:hep-ph/0504231 [hep-ph]].

\bibitem{Alarcon:2022ero}
R.~Alarcon, J.~Alexander, V.~Anastassopoulos, T.~Aoki, R.~Baartman, S.~Bae{\ss}ler, L.~Bartoszek, D.~H.~Beck, F.~Bedeschi and R.~Berger, \textit{et al.}
``Electric dipole moments and the search for new physics,'' Contribution to: Snowmass 2021, [arXiv:2203.08103 [hep-ph]].

\bibitem{Hudson:2011zz}
J.~J.~Hudson, D.~M.~Kara, I.~J.~Smallman, B.~E.~Sauer, M.~R.~Tarbutt and E.~A.~Hinds,
``Improved measurement of the shape of the electron,''
Nature \textbf{473}, 493-496 (2011).

\bibitem{ACME:2013pal}
J.~Baron \textit{et al.} [ACME],
``Order of Magnitude Smaller Limit on the Electric Dipole Moment of the Electron,''
Science \textbf{343}, 269-272 (2014)
[arXiv:1310.7534 [physics.atom-ph]].

\bibitem{ACME:2018yjb}
V.~Andreev \textit{et al.} [ACME],
``Improved limit on the electric dipole moment of the electron,''
Nature \textbf{562}, no.7727, 355-360 (2018).

\bibitem{NLeEDM:2018}
P.~Aggarwal \textit{et al.} [NL-$e$EDM],
``Measuring the electric dipole moment of the electron in BaF,''
[arXiv:1804.10012 [physics.atom-ph]].

\bibitem{Regan:2002ta}
B.~C.~Regan, E.~D.~Commins, C.~J.~Schmidt and D.~DeMille,
``New limit on the electron electric dipole moment,''
Phys. Rev. Lett. \textbf{88}, 071805 (2002).

\bibitem{Cairncross:2017fip}
W.~B.~Cairncross, D.~N.~Gresh, M.~Grau, K.~C.~Cossel, T.~S.~Roussy, Y.~Ni, Y.~Zhou, J.~Ye and E.~A.~Cornell,
``Precision Measurement of the Electron{\textquoteright}s Electric Dipole Moment Using Trapped Molecular Ions,''
Phys. Rev. Lett. \textbf{119}, no.15, 153001 (2017)
[arXiv:1704.07928 [physics.atom-ph]].

\bibitem{Roussy:2022cmp}
T.~S.~Roussy, L.~Caldwell, T.~Wright, W.~B.~Cairncross, Y.~Shagam, K.~B.~Ng, N.~Schlossberger, S.~Y.~Park, A.~Wang and J.~Ye, \textit{et al.}
``An improved bound on the electron{\textquoteright}s electric dipole moment,''
Science \textbf{381}, no.6653, adg4084 (2023)
[arXiv:2212.11841 [physics.atom-ph]].

\bibitem{Ardu:2025rqy}
M.~Ardu and N.~Valori,
``Equivalent electric dipole moment in SMEFT,''
Phys. Rev. D \textbf{113}, no.1, 015035 (2026)
[arXiv:2503.21920 [hep-ph]].

\bibitem{Kozyryev:2017cwq}
I.~Kozyryev and N.~R.~Hutzler,
``Precision Measurement of Time-Reversal Symmetry Violation with Laser-Cooled Polyatomic Molecules,''
Phys. Rev. Lett. \textbf{119}, no.13, 133002 (2017)
[arXiv:1705.11020 [physics.atom-ph]].

\bibitem{Muong-2:2008ebm}
G.~W.~Bennett \textit{et al.} [Muon (g-2)],
``An Improved Limit on the Muon Electric Dipole Moment,''
Phys. Rev. D \textbf{80}, 052008 (2009)
[arXiv:0811.1207 [hep-ex]].

\bibitem{Belle:2021ybo}
K.~Inami \textit{et al.} [Belle],
``An improved search for the electric dipole moment of the $\tau$ lepton,''
JHEP \textbf{04}, 110 (2022)
[arXiv:2108.11543 [hep-ex]].

\bibitem{Abel:2020pzs}
C.~Abel, S.~Afach, N.~J.~Ayres, C.~A.~Baker, G.~Ban, G.~Bison, K.~Bodek, V.~Bondar, M.~Burghoff and E.~Chanel, \textit{et al.}
``Measurement of the Permanent Electric Dipole Moment of the Neutron,''
Phys. Rev. Lett. \textbf{124}, no.8, 081803 (2020)
[arXiv:2001.11966 [hep-ex]].

\bibitem{Graner:2016}
B.~Graner, Y~ Chen, E.~G.~Lindahl, B.~R.~Heckel,
``Reduced Limit on the Permanent Electric Dipole Moment of ${}^{199}$Hg,''
Phys. Rev. Lett. \textbf{116}, 161601 (2016)
[arXiv:1601.04339 [physics.atom-ph]].

\bibitem{Anastassopoulos:2015ura}
V.~Anastassopoulos, S.~Andrianov, R.~Baartman, M.~Bai, S.~Baessler, J.~Benante, M.~Berz, M.~Blaskiewicz, T.~Bowcock and K.~Brown, \textit{et al.}
``A Storage Ring Experiment to Detect a Proton Electric Dipole Moment,''
Rev. Sci. Instrum. \textbf{87}, no.11, 115116 (2016)
[arXiv:1502.04317 [physics.acc-ph]].

\bibitem{Haciomeroglu:2018nre}
S.~Haciomeroglu and Y.~K.~Semertzidis,
``Hybrid ring design in the storage-ring proton electric dipole moment experiment,''
Phys. Rev. Accel. Beams \textbf{22}, no.3, 034001 (2019)
[arXiv:1806.09319 [physics.acc-ph]].

\bibitem{1865072}
V.~Anastassopoulos \textit{et al.} [Storage Ring EDM],
``A Proposal to Measure the Proton Electric Dipole Moment with $10^{-29}$ $ e \cdot$cm Sensitivity,''
\url{https://inspirehep.net/files/fedd912e77ee5f1defd288d2ea8f8aeb}, \url{https://www.bnl.gov/edm/index.html}.

\bibitem{Omarov:2020kws}
Z.~Omarov, H.~Davoudiasl, S.~Haciomeroglu, V.~Lebedev, W.~M.~Morse, Y.~K.~Semertzidis, A.~J.~Silenko, E.~J.~Stephenson and R.~Suleiman,
``Comprehensive symmetric-hybrid ring design for a proton EDM experiment at below 10-29e{\textperiodcentered}cm,''
Phys. Rev. D \textbf{105}, no.3, 032001 (2022)
[arXiv:2007.10332 [physics.acc-ph]].

\bibitem{CPEDM:2019nwp}
F.~Abusaif \textit{et al.} [CPEDM],
``Storage ring to search for electric dipole moments of charged particles: Feasibility study,'' CERN, 2021,
[arXiv:1912.07881 [hep-ex]].

\bibitem{Yamanaka:2018dwa}
N.~Yamanaka,
``Electric dipole moment of light nuclei {\ensuremath{-}} $^6$Li, $^7$Li, $^9$Be, $^{11}$B, and $^{13}$C {\ensuremath{-}},''
Hyperfine Interact. \textbf{239}, 35 (2018)
[arXiv:1805.05982 [nucl-th]].

\bibitem{Liu:2007en}
C.~P.~Liu and J.~Engel,
``Schiff screening of relativistic nucleon electric-dipole moments by electrons,''
Phys. Rev. C \textbf{76}, 028501 (2007)
[arXiv:0705.1981 [nucl-th]].

\bibitem{Khriplovich:1998zq}
I.~B.~Khriplovich,
``Feasibility of search for nuclear electric dipole moments at ion storage rings,''
Phys. Lett. B \textbf{444}, 98-102 (1998)
[arXiv:hep-ph/9809336 [hep-ph]].

\bibitem{JEDI:2016swi}
G.~Guidoboni \textit{et al.} [JEDI],
``How to Reach a Thousand-Second in-Plane Polarization Lifetime with 0.97-GeV/c Deuterons in a Storage Ring,''
Phys. Rev. Lett. \textbf{117}, no.5, 054801 (2016).

\bibitem{Peccei:1977hh}
R.~D.~Peccei and H.~R.~Quinn,
``CP Conservation in the Presence of Instantons,''
Phys. Rev. Lett. \textbf{38}, 1440-1443 (1977).

\bibitem{Peccei:1977ur}
R.~D.~Peccei and H.~R.~Quinn,
``Constraints Imposed by CP Conservation in the Presence of Instantons,''
Phys. Rev. D \textbf{16}, 1791-1797 (1977)

\bibitem{Irastorza:2021tdu}
I.~G.~Irastorza,
``An introduction to axions and their detection,''
SciPost Phys. Lect. Notes \textbf{45}, 1 (2022)
[arXiv:2109.07376 [hep-ph]].

\bibitem{Weinberg:1977ma}
S.~Weinberg,
``A New Light Boson?,''
Phys. Rev. Lett. \textbf{40}, 223-226 (1978).

\bibitem{Wilczek:1977pj}
F.~Wilczek,
``Problem of Strong  $P$  and  $T$  Invariance in the Presence of Instantons,''
Phys. Rev. Lett. \textbf{40}, 279-282 (1978).

\bibitem{Wilczek:1991jgb}
F.~Wilczek,
``The Birth of Axions,'' 
Current Contents \textbf{16}, 8-9 (1991).

\bibitem{Peccei:2006as}
R.~D.~Peccei,
``The Strong CP problem and axions,''
Lect. Notes Phys. \textbf{741}, 3-17 (2008)
[arXiv:hep-ph/0607268 [hep-ph]].

\bibitem{DiLuzio:2020wdo}
L.~Di Luzio, M.~Giannotti, E.~Nardi and L.~Visinelli,
``The landscape of QCD axion models,''
Phys. Rept. \textbf{870}, 1-117 (2020)
[arXiv:2003.01100 [hep-ph]].

\bibitem{Svrcek:2006yi}
P.~Svrcek and E.~Witten,
``Axions In String Theory,''
JHEP \textbf{06}, 051 (2006)
[arXiv:hep-th/0605206 [hep-th]].

\bibitem{Arvanitaki:2009fg}
A.~Arvanitaki, S.~Dimopoulos, S.~Dubovsky, N.~Kaloper and J.~March-Russell,
``String Axiverse,''
Phys. Rev. D \textbf{81}, 123530 (2010)
[arXiv:0905.4720 [hep-th]].

\bibitem{Kim:1979if}
J.~E.~Kim,
``Weak Interaction Singlet and Strong CP Invariance,''
Phys. Rev. Lett. \textbf{43}, 103 (1979).

\bibitem{Shifman:1979if}
M.~A.~Shifman, A.~I.~Vainshtein and V.~I.~Zakharov,
``Can Confinement Ensure Natural CP Invariance of Strong Interactions?,''
Nucl. Phys. B \textbf{166}, 493-506 (1980).

\bibitem{Dine:1981rt}
M.~Dine, W.~Fischler and M.~Srednicki,
``A Simple Solution to the Strong CP Problem with a Harmless Axion,''
Phys. Lett. B \textbf{104}, 199-202 (1981).

\bibitem{Zhitnitsky:1980tq}
A.~R.~Zhitnitsky,
``On Possible Suppression of the Axion Hadron Interactions (in Russian),''
Sov. J. Nucl. Phys. \textbf{31}, 260 (1980).

\bibitem{Irastorza:2018dyq}
I.~G.~Irastorza and J.~Redondo,
``New experimental approaches in the search for axion-like particles,''
Prog. Part. Nucl. Phys. \textbf{102}, 89-159 (2018)
[arXiv:1801.08127 [hep-ph]].

\bibitem{Derbenev:1975}
Y.~S.~Derbenev and A.~Kondratenko,
``Acceleration of polarized particles,''
Doklady Akademii Nauk SSSR (in Russian), \textbf{223}, no.4, 830 (1975).

\bibitem{Derbenev:2024jyn}
Y.~S.~Derbenev,
``The Twisted Spin Synchrotron,'' (1996) [arXiv:2404.00073 [physics.acc-ph]].

\bibitem{Talman:2015pma}
R.~M.~Talman and J.~D.~Talman,
``Electric dipole moment planning with a resurrected BNL Alternating Gradient Synchrotron electron analog ring,''
Phys. Rev. ST Accel. Beams \textbf{18}, no.7, 074004 (2015)
[arXiv:1503.08494 [physics.acc-ph]].

\bibitem{Agrawal:2022lsp}
P.~Agrawal, M.~Nee and M.~Reig,
``Axion couplings in grand unified theories,''
JHEP \textbf{10}, 141 (2022)
[arXiv:2206.07053 [hep-ph]].

\bibitem{Agrawal:2024ejr}
P.~Agrawal, M.~Nee and M.~Reig,
``Axion couplings in heterotic string theory,''
JHEP \textbf{02}, 188 (2025)
[arXiv:2410.03820 [hep-ph]].

\bibitem{Reig:2025dqb}
M.~Reig and T.~Weigand,
``Testing the heterotic string with the axion-photon coupling,''
JHEP \textbf{01}, 006 (2026)
[arXiv:2509.08042 [hep-th]].

\bibitem{Fukuyama:2013ioa}
T.~Fukuyama and A.~J.~Silenko,
``Derivation of Generalized Thomas-Bargmann-Michel-Telegdi Equation for a Particle with Electric Dipole Moment,''
Int. J. Mod. Phys. A \textbf{28}, 1350147 (2013)
[arXiv:1308.1580 [hep-ph]].

\bibitem{Thomas:1927yu}
L.~H.~Thomas,
``The Kinematics of an electron with an axis,''
Phil. Mag. Ser. 7 \textbf{3}, 1-21 (1927).

\bibitem{Filatov:2020ygb}
Y.~N.~Filatov, A.~M.~Kondratenko, M.~A.~Kondratenko, Y.~S.~Derbenev and V.~S.~Morozov,
``Transparent Spin Method for Spin Control of Hadron Beams in Colliders,''
Phys. Rev. Lett. \textbf{124}, no.19, 194801 (2020)
[arXiv:2003.11469 [physics.acc-ph]].

\bibitem{Kondratenko:2019ubq}
A.~Kondratenko, M.~Kondratenko, Y.~Filatov, A.~Kovalenko, Y.~Derbenev and V.~Morozov,
``Feasibility of measuring EDM in spin transparent colliders,''
EPJ Web Conf. \textbf{204}, 10013 (2019).

\bibitem{Flanz:1985lnw}
J.~B.~Flanz and C.~P.~Sargent,
``Operation of an isochronous beam recirculation system,''
Nucl. Instrum. Meth. A \textbf{241}, 325-333 (1985).

\bibitem{Farley:2003wt}
F.~J.~M.~Farley, K.~Jungmann, J.~P.~Miller, W.~M.~Morse, Y.~F.~Orlov, B.~L.~Roberts, Y.~K.~Semertzidis, A.~Silenko and E.~J.~Stephenson,
``A New method of measuring electric dipole moments in storage rings,''
Phys. Rev. Lett. \textbf{93}, 052001 (2004)
[arXiv:hep-ex/0307006 [hep-ex]].

\bibitem{Kim:2021pld}
O.~Kim and Y.~K.~Semertzidis,
``New method of probing an oscillating EDM induced by axionlike dark matter using an rf Wien filter in storage rings,''
Phys. Rev. D \textbf{104}, no.9, 096006 (2021)
[arXiv:2105.06655 [hep-ph]].

\bibitem{Carli:2022ftv}
C.~Carli and M.~Haj Tahar,
``Geometric phase effect study in electric dipole moment rings,''
Phys. Rev. Accel. Beams \textbf{25}, no.6, 064001 (2022).

\bibitem{Murphy:1996yt}
J.~B.~Murphy, S.~Krinsky and R.~L.~Gluckstern,
``Longitudinal wake field for an electron moving on a circular orbit,''
Part. Accel. \textbf{57}, 9-64 (1997) BNL-63090, 
\url{https://inspirehep.net/files/1afd5dcc4e791d233fb0ffb5a2f270a4}.

\bibitem{Warnock:1988sb}
R.~L.~Warnock and P.~L.~Morton,
Part. Accel. \textbf{25}, 113 (1990) SLAC-PUB-4562,
\url{https://slac.stanford.edu/pubs/slacpubs/4500/slac-pub-4562.pdf}.

\bibitem{Brantjes:2012zz}
N.~P.~M.~Brantjes, V.~Dzhordzhadze, R.~Gebel, F.~Gonnella, F.~E.~Gray, D.~J.~van der Hoek, A.~Imig, W.~L.~Kruithof, D.~M.~Lazarus and A.~Lehrach, \textit{et al.}
``Correcting systematic errors in high-sensitivity deuteron polarization measurements,''
Nucl. Instrum. Meth. A \textbf{664}, 49-64 (2012).

\bibitem{Graham:2020kai}
P.~W.~Graham, S.~Haciomeroglu, D.~E.~Kaplan, Z.~Omarov, S.~Rajendran and Y.~K.~Semertzidis,
``Storage ring probes of dark matter and dark energy,''
Phys. Rev. D \textbf{103}, no.5, 055010 (2021)
[arXiv:2005.11867 [hep-ph]].

\bibitem{Fan:2023hci}
X.~Fan and M.~Reig,
``New bounds and future prospects for axion force searches at Penning trap experiments,''
[arXiv:2310.18797 [hep-ph]].

\bibitem{Agrawal:2022wjm}
P.~Agrawal, D.~E.~Kaplan, O.~Kim, S.~Rajendran and M.~Reig,
``Searching for axion forces with precision precession in storage rings,''
Phys. Rev. D \textbf{108}, no.1, 015017 (2023)
[arXiv:2210.17547 [hep-ph]].

\bibitem{Agrawal:2023lmw}
P.~Agrawal, N.~R.~Hutzler, D.~E.~Kaplan, S.~Rajendran and M.~Reig,
``Searching for axion forces with spin precession in atoms and molecules,''
JHEP \textbf{07}, 133 (2024)
[arXiv:2309.10023 [hep-ph]].

\bibitem{Jakubassa:2026}
D.~Jakubassa,
``The Sherman function and its radiative corrections for elastic positron-nucleus scattering,''
\url{https://indico.jlab.org/event/964/contributions/18030/attachments/13893/22471/DJA-JLab26.pdf}, to appear in LEEPP2026 conference proceedings.

\bibitem{Aghababaei:2024hcb}
S.~Aghababaei,
``Constraints on Lorentz violation parameter through electric dipole moments,''
Eur. Phys. J. C \textbf{84}, no.2, 173 (2024).

\bibitem{Bluhm:1998wf}
R.~Bluhm, V.~A.~Kostelecky and N.~Russell,
``Testing CPT and Lorentz symmetry with electrons and positrons in Penning traps,''
AIP Conf. Proc. \textbf{457}, no.1, 138 (1999)
[arXiv:hep-ph/9810287 [hep-ph]].

\bibitem{Ding:2026aac}
Y.~Ding,
``Lorentz and CPT Tests in Neutron and Storage-Ring EDM Experiments,''
[arXiv:2601.08899 [hep-ph]].

\bibitem{Derbenev:2021}
Ya~S.~Derbenev, Yu.~N.~Filatov, A.~M.~Kondratenko, M.~A.~Kondratenko and V.~S.~Morozov,
``Siberian snakes, figure-8 and spin transparency techniques for high precision experiments with polarized hadron beams in colliders,''
Symmetry \textbf{13}, 398 (2021), \url{https://www.mdpi.com/2073-8994/13/3/398}.

\end{thebibliography}
